\documentclass[pdflatex,sn-nature]{sn-jnl}
\setcitestyle{numbers,super,open={},close={}}

\usepackage{graphicx}%
\usepackage{multirow}%
\usepackage{amsmath,amssymb,amsfonts}%
\usepackage{amsthm}%
\usepackage{mathrsfs}%
\usepackage[title]{appendix}%
\usepackage{xcolor}%
\usepackage{textcomp}%
\usepackage{manyfoot}%
\usepackage{booktabs}%
\usepackage{algorithm}%
\usepackage{algorithmicx}%
\usepackage{algpseudocode}%
\usepackage{listings}%
\usepackage{pifont}
\usepackage{rotating}
\usepackage[switch]{lineno}

\theoremstyle{thmstyleone}%
\theoremstyle{thmstyletwo}%

\theoremstyle{thmstylethree}%

\newenvironment{sidewaystableplain}[1][]%
{\begin{sidewaystableorg}[#1]\begin{center}}
{\end{center}\end{sidewaystableorg}}

\newenvironment{plaintable}[1][]%
{\begin{tableorg}[#1]\begin{center}}
{\end{center}\end{tableorg}}

\begin{document}


\title[Article Title]{Task-adaptive anonymisation enables privacy-preserving sharing of clinical speech data}



\author[1]{\fnm{Ziyun} \sur{Cui}}
\equalcont{These authors contributed equally to this work.}
\author[1]{\fnm{Haoyu} \sur{Niu}}
\equalcont{These authors contributed equally to this work.}
\author[2]{\fnm{Wen} \sur{Wu}}
\equalcont{These authors contributed equally to this work.}
\author[3]{\fnm{Hao} \sur{Sun}}
\author[4]{\fnm{Sihan} \sur{Zhang}}
\author[5]{\fnm{Chang} \sur{Lei}}
\author[6]{\fnm{Yun} \sur{Zhang}}
\author[7,8,9]{\fnm{Chunyun} \sur{Sun}}
\author[7,8,9]{\fnm{Yang} \sur{Luo}}
\author[10]{\fnm{Liangliang} \sur{Cao}}
\author[1]{\fnm{Ji} \sur{Wu}}
\author*[5]{\fnm{Runsen} \sur{Chen}}\email{runsenchen@tsinghua.edu.cn}
\author*[6,11]{\fnm{Lian} \sur{Zhou}}\email{zlpumch02@163.com}
\author*[7,8,9]{\fnm{Yongsheng} \sur{Tong}}\email{timystong@pku.org.cn}
\author*[1,2,12]{\fnm{Chao} \sur{Zhang}}\email{cz277@tsinghua.edu.cn}

\affil[1]{\orgname{Department of Electronic Engineering, Tsinghua University}, \orgaddress{\city{Beijing} 100084, \country{China}}}
\affil[2]{\orgname{Shanghai Artificial Intelligence Laboratory}, \orgaddress{\city{Shanghai} 200032, \country{China}}}
\affil[3]{\orgname{School of Computer and Communication Engineering, University of Science and Technology Beijing}, \orgaddress{\city{Beijing} 100083, \country{China}}}
\affil[4]{\orgname{Beijing Tsinghua Changgung Hospital, School of Clinical Medicine, Tsinghua Medicine, Tsinghua University}, \orgaddress{\city{Beijing} 100084, \country{China}}}
\affil[5]{\orgname{Vanke School of Public Health, Tsinghua University}, \orgaddress{\city{Beijing} 100084, \country{China}}}
\affil[6]{\orgname{Peking Union Medical College Hospital}, \orgaddress{\city{Beijing} 100730, \country{China}}}
\affil[7]{\orgname{Peking University Huilongguan Clinical Medical School}, \orgaddress{\city{Beijing} 100096, \country{China}}}
\affil[8]{\orgname{Department of Clinical Psychology, Beijing Huilongguan Hospital, Capital Medical University}, \orgaddress{\city{Beijing} 100096, \country{China}}}
\affil[9]{\orgname{WHO Collaborating Center for Research and Training in Suicide Prevention}, \orgaddress{\city{Beijing} 100096, \country{China}}}
\affil[10]{\orgname{Department of Data Science and Artificial Intelligence, The Hong Kong Polytechnic University}, \orgaddress{\city{Hong Kong (SAR)}, \country{China}}}
\affil[11]{\orgname{Chinese Academy of Medical Sciences}, \orgaddress{\city{Beijing} 100730, \country{China}}}
\affil[12]{\orgname{Department of Speech, Hearing and Phonetic Sciences, University College London (UCL)}, \orgaddress{\city{London} WC1N 1PF, \country{United Kingdom}}}








\abstract{
Clinical speech is increasingly used for disease screening and clinical assessment, but directly distributing speech recordings poses privacy risks because they contain both speaker-identifying voiceprints and sensitive spoken content. Existing speech anonymisation approaches are primarily developed for general speech applications or task-specific clinical speech and overlook that different clinical tasks rely on distinct speech characteristics, necessitating task-adaptive anonymisation strategies to protect patient privacy while preserving clinical utility.
Here we introduce {CASA}, a \textbf{c}linically \textbf{a}daptive \textbf{s}peech \textbf{a}nonymisation agent that transforms anonymisation from a fixed processing procedure into a task-guided optimisation process, dynamically tailoring voiceprint and content anonymisation to task-specific privacy and clinical utility requirements. Through recursive self-improvement (RSI), CASA iteratively evaluates and refines anonymisation strategies against task-specific privacy and clinical utility requirements, while learning from these outcomes to guide more efficient planning in subsequent clinical tasks.
CASA is validated across 14 clinical tasks encompassing five languages, diverse elicitation paradigms, and mental health, cognitive and voice disorders.
CASA preserved disease-related classification performance while meeting speaker-anonymisation requirements across all 14 clinical tasks (14/14), versus only 8/14 for the strongest fixed method.
Assessments of resistance to speaker re-identification by external attackers and identification and redaction of sensitive personal information in clinical speech further demonstrated CASA's dual protection of speaker identity and sensitive spoken content.
Beyond anonymisation effectiveness, CASA's ability to adapt to new clinical tasks was enhanced by accumulated experience of RSI, reducing the number of strategy evaluations needed to meet privacy and clinical utility requirements by 31.3\%.
Clinical expert assessments further supported the preservation of clinically relevant information and the acceptability of anonymised speech for controlled research sharing. 

}

\maketitle


Speech arises from the coordinated interaction of auditory perception, neural processing, motor planning and the respiratory, phonatory and articulatory systems, producing a complex behavioural signal that reflects both physiological and neurological function~\cite{denes2015speech}. Because speech can be collected remotely and non-invasively, it has emerged as a promising digital biomarker for a broad range of clinical applications~\cite{fagherazzi2021voice, sara2023guess, de2025clinical}. Previous studies have demonstrated the utility of speech-based analysis for mental health assessment~\cite{cummins2015review, ringeval2019avec, flanagan2021using, wu2023self}, cognitive impairment screening~\cite{luz2021alzheimer, ivanova2022discriminating, cui2023transferring}, and speech disorder evaluation~\cite{putzer2008saarbruencken, hasegawa2024community,assessment_is26}. However, speech simultaneously encodes highly identifiable information, including speaker identity and spoken linguistic content, creating substantial privacy risks for the collection, sharing and analysis of clinical speech data~\cite{sara2023guess, diaz2025navigating, malo2026using}. Effective anonymisation is therefore essential for enabling the safe sharing and analysis of clinical speech data while preserving information required for downstream clinical assessment.

A major challenge for clinical speech anonymisation is that different clinical tasks rely on distinct speech characteristics for assessment, including acoustic, phonetic, paralinguistic and linguistic information~\cite{diaz2025navigating}. At the same time, different elicitation paradigms expose different forms of privacy-sensitive information. For example, reading speech may primarily require suppression of speaker-identifiable vocal characteristics~\cite{tayebi2023effect}, whereas spontaneous autobiographical speech may additionally contain sensitive semantic content~\cite{aggazzotti2026content}. Consequently, clinical speech anonymisation is unlikely to admit a one-size-fits-all solution. The goal is not simply to maximise anonymisation strength, but to achieve sufficient privacy protection while preserving the speech cues required for a given clinical task.

Practical clinical settings therefore require scalable speech anonymisation methods that can operate across diverse diseases and elicitation paradigms. Existing anonymisation approaches, however, remain largely fragmented between general-purpose speech anonymisation and disease-specific clinical solutions. Methods developed for general speech applications~\cite{tomashenko2022voiceprivacy} primarily focus on suppressing speaker identity and often inadvertently degrade clinically relevant speech information, while disease-specific anonymisation approaches~\cite{ravuri2022preserving,tayebi2024addressing,cui2026speaker} are typically designed for individual disorders or narrowly defined tasks, limiting their applicability across heterogeneous clinical settings. 
To our knowledge, no existing framework adapts both voiceprint and spoken-content anonymisation to individual clinical tasks across diseases, languages and elicitation paradigms.

Here we present \textbf{CASA}, a \textbf{c}linically \textbf{a}daptive \textbf{s}peech \textbf{a}nonymisation agent that reformulates clinical speech anonymisation as a task-adaptive privacy–utility optimisation problem rather than a fixed preprocessing procedure. 
Given a natural-language description of clinical task properties and corresponding privacy requirements, CASA autonomously devises voiceprint and content anonymisation strategies to protect patient privacy while preserving disease-relevant speech information. Its recursive self-improvement mechanism iteratively evaluates and refines these strategies based on privacy–utility feedback for the current task, while distilling successful strategies and failure cases into memory to inform subsequent tasks. 
This creates a continuous learning cycle in which each task both draws on and contributes to CASA's growing experience, progressively refining its strategy planning. In our evaluation, this experience-guided planning reduced the number of iterations required to meet the privacy–utility requirements by 31.3\%.
We validate CASA on 11 clinical speech datasets encompassing 14 clinical tasks, five languages and diverse mental health, cognitive and voice disorders. 
CASA satisfied the joint speaker-privacy and clinical-utility requirements across all 14 tasks, compared with at most eight for any individual fixed anonymisation method.
The preservation of clinically relevant information and the acceptability of anonymised recordings for controlled research sharing are further supported by clinical expert evaluation.
We envisage that this approach could facilitate broader clinical applications of speech-based assessment and privacy-preserving data sharing for multi-centre collaborative research.

\begin{figure}[ht!]
\centering
\includegraphics[width=0.98\textwidth]{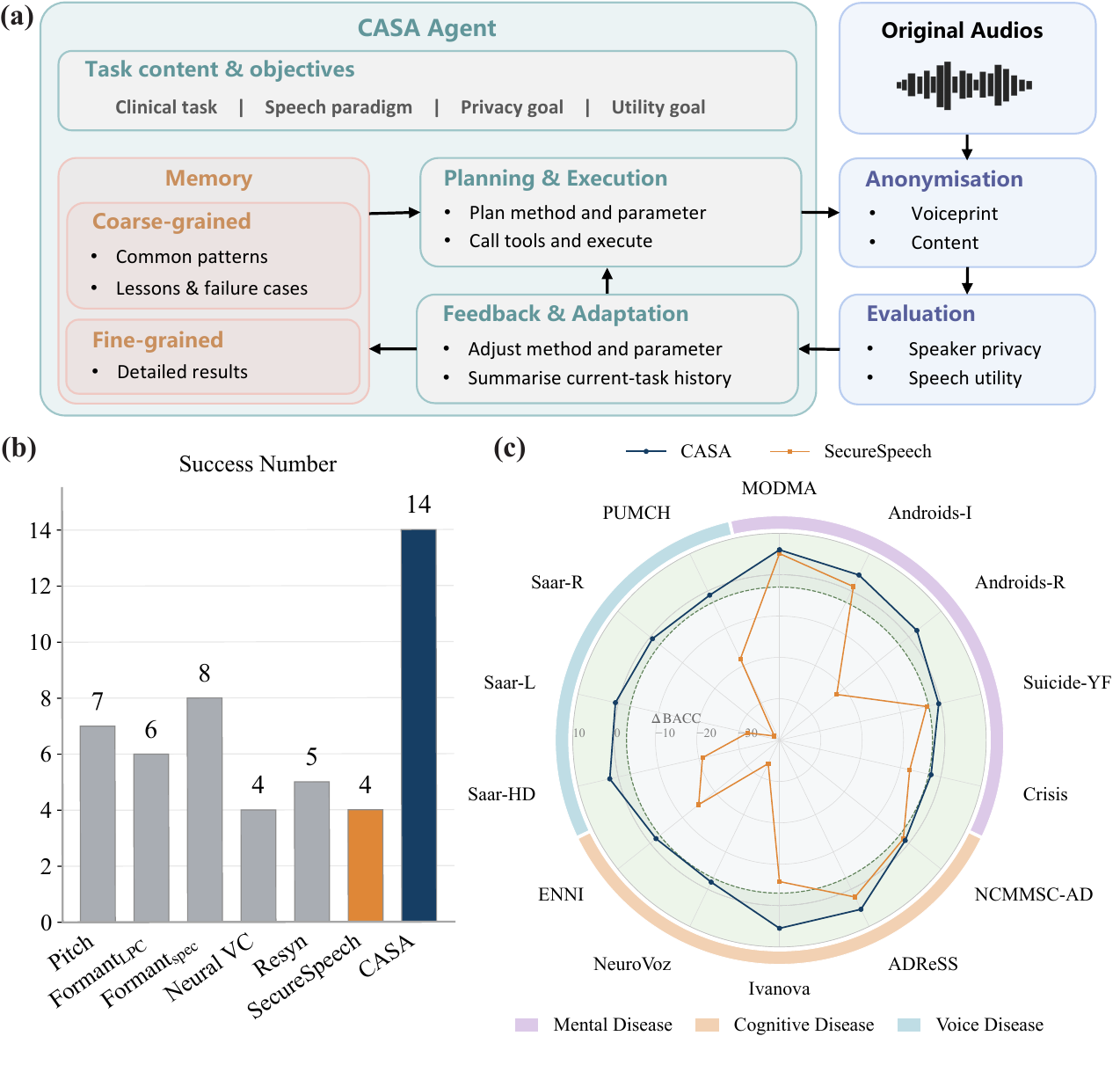}
\caption{\textbf{$|$ Overview of CASA and comparison with baselines.} 
a, Architecture of the CASA agent. An LLM-based controller receives the clinical task, speech paradigm and privacy–utility requirements, plans an anonymisation strategy and invokes tools for voiceprint and content anonymisation. Evaluation feedback guides adjustment to methods and parameters. Cross-task memory retains coarse-grained summaries and fine-grained experimental records to guide subsequent planning. 
b, Number of datasets or task subsets meeting all applicable privacy and clinical-utility criteria, comparing five individual methods used as fixed strategies (grey), a SecureSpeech pipeline (orange) and CASA (blue). Success is defined as that the anonymised speech achieve an EER of at least 20\%, sensitive-content localisation F1 of at least 0.75 when required, and a downstream BACC decrease of less than three percentage points relative to original speech. 
c, Delta BACC in percentage points after anonymisation with CASA and SecureSpeech. The dashed green circle marks the -3-percentage-point boundary, and the green shaded outer region indicates changes satisfying the utility requirement. The datasets are grouped into mental health, cognitive and voice disorders.
}
\label{fig:main}
\end{figure}

\section*{Results}
\label{sec: result}

We conducted a series of evaluations to establish whether CASA can reconcile patient privacy with clinical utility across heterogeneous speech-based clinical tasks. We first compared CASA with other anonymisation methods to assess its ability to satisfy task-specific speaker-privacy and clinical-utility requirements across diverse tasks, and evaluated its protection against external speaker re-identification attackers. We then examined how different voiceprint anonymisation methods affect speaker privacy and the preservation of speech characteristics, investigating why task-specific strategy selection is necessary. We also evaluated CASA's performance in identifying and protecting sensitive spoken content, as well as its ability to improve the efficiency of task-adaptive anonymisation strategy planning through recursive self-improvement. Finally, clinical expert assessments examined whether anonymised recordings retained clinically relevant information and were considered acceptable for controlled research sharing.

\subsection*{CASA satisfies privacy–utility requirements across diverse clinical speech tasks}

We first investigated whether CASA could satisfy joint speaker-privacy and clinical-utility requirements across heterogeneous clinical tasks where fixed anonymisation strategies fall short. Unlike fixed pipelines, CASA determines voiceprint and content anonymisation strategies according to task-specific requirements and iteratively adjusts them based on evaluation feedback (Fig.~\ref{fig:main}a). It integrates five voiceprint anonymisation methods -- pitch shifting (\texttt{Pitch}), linear predictive coding (LPC)-based formant transformation (\texttt{Formant$_\text{LPC}$}), spectral formant scaling (\texttt{Formant$_\text{spec}$}),  neural voice conversion (\texttt{Neural VC}) and transcription–resynthesis (\texttt{Resyn})-- alongside large language model (LLM)-based localisation and masking of sensitive spoken content. These components provide the available anonymisation operations, while the agent determines when and how to apply them for a given clinical task.

\begin{plaintable}[t]
\centering
\caption{\textbf{$|$ Clinical speech datasets applied for validation, spanning mental health disorders, cognitive disorders, and voice disorders.} Androids-Corpus was evaluated separately for interview and reading speech (Androids-I and Androids-R), and Saarbrücken was split into three disease-versus-control subsets (Saar-HD, Saar-L and Saar-R), resulting in 14 evaluation tasks/subsets. Disease type, language, number of subjects, whether recording is multi-speaker (Multi-Spk), data source, and downstream evaluation tasks, are listed. Downstream tasks include binary or multi-class classification.}
\label{tab: dataset}
\resizebox{1\linewidth}{!}{
\begin{tabular}{@{}ccccccc@{}}
\toprule
Dataset                      & Disease    & Language & N-Subjects & Multi-Spk & Source              & Downstream   \\ \midrule
MODMA                        & Depression & Chinese  & 52         & \ding{55} & \cite{cai2022multi} & Binary       \\
Androids-Corpus              & Depression & Italian  & 118        & \ding{51} & \cite{tao2023androids}        & Binary       \\ 
Suicide-YF                   & Suicide risk & Chinese & 1223 & \ding{55} & In-house            &  Binary  \\ 
Crisis                       &  Psychological crisis & Chinese  & 154        & \ding{51} & In-house            & 3-class      \\ \midrule
NCMMSC-AD                   & Alzheimer's Disease  & Chinese  & 175        & \ding{55} & \cite{chen2023raw} & 3-class      \\
ADReSS                       & Alzheimer's Disease  & English  & 156        & \ding{55} & \cite{luz2021alzheimer}       & Binary       \\
Ivanova                      & Alzheimer's Disease & Spanish & 357 & \ding{55} & \cite{ivanova2022discriminating} & 3-class \\
NeuroVoz                     & Parkinson's Disease  & Spanish  & 112        & \ding{55} & \cite{mendes2024neurovoz}  & Binary \\
ENNI                         & Language impairment & English  & 352        & \ding{51} & \cite{schneider2006storytelling} & Binary \\ \midrule
\multirow{3}{*}{Saarbrücken} & Hyperfunktionelle Dysphonie         & \multirow{3}{*}{German}   & 424        & \ding{55} & \multirow{3}{*}{\cite{putzer2008saarbruencken}} & Binary            \\
                             & Laryngitis &  & 280        & \ding{55} &                     & Binary             \\
                             & Rekurrensparese   &    & 424        & \ding{55} &                     & Binary             \\
PUMCH                        & Cleft-palate & Chinese & 244 & \ding{51} & In-house            & Binary \\
\bottomrule
\end{tabular}
}
\end{plaintable}

To assess the benefit of task-adaptive anonymisation strategy, we compared CASA with each of the five voiceprint anonymisation methods applied as a fixed strategy. We also included a multilingual implementation of SecureSpeech~\cite{hui2025securespeech}, which combines voiceprint and content anonymisation in a fixed pipeline, as a complete-system baseline. The evaluation covered 11 clinical speech datasets comprising 14 tasks or subsets across five languages, spanning mental health, cognitive and voice disorders and including both single-speaker and multi-speaker recordings (Table~\ref{tab: dataset}). This heterogeneous setting allowed us to examine whether anonymisation strategies could satisfy the same privacy and utility requirements across different clinical contexts. 

We defined successful anonymisation using three simultaneous criteria: (i) a speaker-verification  equal error rate (EER) of at least 20\%, indicating that the specified speaker-privacy threshold was met; (ii) a F1 score at least 0.75 for localisation of sensitive spoken content when present and required by the task, indicating that the content privacy requirement was met; and (iii)  a reduction in downstream balanced accuracy (BACC) of less than three percentage points relative to original speech, indicating preservation of disease-related classification performance. 

As shown in Fig.~\ref{fig:main}b, CASA succeeded on all 14 clinical tasks (100\%) under these criteria, whereas the best-performing individual fixed method succeeded on only 8 tasks and SecureSpeech on 4 tasks. 
Notably, the Crisis and PUMCH datasets contain sensitive spoken information requiring both voiceprint and content anonymisation. The five individual voiceprint anonymisation methods could not fulfil this additional requirement, whereas CASA and SecureSpeech both incorporate content anonymisation, enabling a comparison of the two complete systems. As shown in Fig.~\ref{fig:main}c, CASA achieved higher BACC than SecureSpeech on all tasks while maintaining the classification performance decline within the specified tolerance across all tasks. 

We further assessed CASA's voiceprint protection with two external attackers: (i) a semi-informed attacker trained on anonymised speech following the VoicePrivacy training recipe~\cite{tomashenko2026third}, which assesses whether an attacker informed prior knowledge of the speech transformation pipeline can successfully re-identify speakers, and (ii) a speech large language model (SpeechLLM) attacker based on SALMONN-2-8B~\cite{yang2026salmonn} that directly judges whether two speech segments originate from the same speaker. 
The semi-informed attacker yielded a mean EER of 0.338 across the 14 clinical tasks, with all meeting the privacy threshold considered in VoicePrivacy 2024. The SpeechLLM attacker yielded a mean verification error of 0.441, calculated as one minus accuracy of same-speaker versus different-speaker classification. These results demonstrate CASA’s broad resistance to speaker re-identification across diverse clinical speech tasks.

Overall, task-adaptive strategy enables CASA to satisfy joint speaker-privacy and clinical-utility requirements across a wider range of clinical tasks. We next investigate how differences in disease-relevant speech characteristics and privacy-sensitive content across clinical tasks shape the requirements for voiceprint and content anonymisation, respectively.

\subsection*{Anonymisation of speaker voiceprint}
\label{sec: voiceprint result}

\begin{figure}
    \centering
    \includegraphics[width=0.9\linewidth]{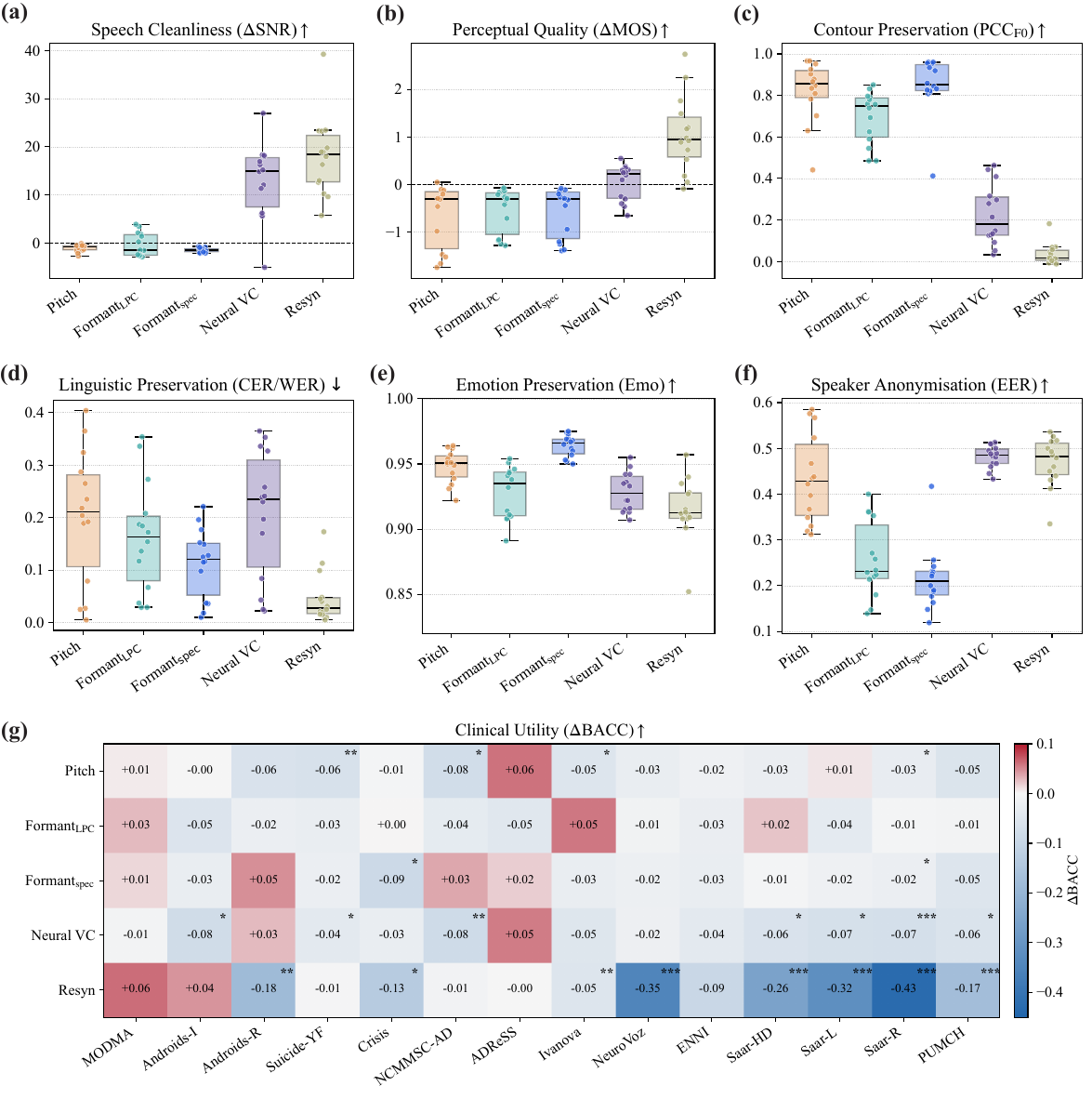}
    \caption{\textbf{$|$ Voiceprint anonymisation evaluation across methods and datasets.} Boxplots show six evaluation metrics across datasets for each voiceprint anonymisation method: $\Delta \text{SNR}$ (change in SNR, anonymised vs. original), $\Delta \text{MOS}$, PCC$_\text{F0}$, CER/WER, Emo, and EER. Each dot represents a single subset/task. Box plots show the median (centre line), the first and third quartiles (box bounds), and whiskers extending to 1.5× the interquartile range. The dashed horizontal line in $\Delta \text{SNR}$ and $\Delta \text{MOS}$ at zero indicates no change relative to the original audio. The heatmap shows $\Delta \text{BACC}$. Negative values (blue) indicate performance degradation and positive values (red) indicate improvement.
    Statistical significance was assessed using one-sided exact binomial tests on discordant paired subject-level predictions. No adjustment was made for multiple comparisons. Exact $P$ values are provided in Supplementary Tables~\ref{tab: result1}--\ref{tab: result3}. Asterisks denote significance levels: $*~P < 0.05$, $**~P < 0.01$, $***~P < 0.001$.
    }
    \label{fig:metrics}
\end{figure}

To systematically understand the benefit of task-adaptive design, we examined how different voiceprint anonymisation methods affected different speech characteristics and downstream clinical performance. Beyond EER and BACC introduced above, our multi-dimensional evaluation further includes signal cleanliness (\textit{i.e.}, signal-to-noise ratio, SNR), perceptual quality (\textit{i.e.}, mean opinion score, MOS), preservation of paralinguistic information (\textit{i.e.}, Pearson correlation coefficient of fundamental frequency (F0), PCC$_\text{F0}$, and emotion similarity, Emo), and preservation of linguistic information (\textit{i.e.}, character and word error rates, CER/WER). Full results are provided in Supplementary Tables~\ref{tab: result1}--\ref{tab: result3}.

\textbf{Method-dependent differences in speaker traceability and speech information preservation.} Results across datasets reveal consistent method-dependent patterns in speech quality (Fig.~\ref{fig:metrics}a,b), speech attributes preservation (Fig.~\ref{fig:metrics}c-e), and speaker anonymisation (Fig.~\ref{fig:metrics}f). 
Signal-processing-based methods (\texttt{Pitch}, \texttt{Formant$_\text{spec}$}, \texttt{Formant$_\text{LPC}$}) generally reduced  perceptual quality while preserving pitch contours more effectively than neural voice conversion (\texttt{Neural VC}) and transcription--resynthesis (\texttt{Resyn}). 
Among these methods, \texttt{Formant$_\text{spec}$} showed the strongest overall preservation of linguistic and emotional information, but also yielded the lowest EER, indicating that preservation of speech characteristics came at the cost of weaker speaker anonymisation. 
\texttt{Pitch} provided stronger anonymisation, with EER values generally above 0.4, but introduced greater transcription errors.
In contrast, \texttt{Neural VC} and \texttt{Resyn} exhibited a different trade-off profile. 
\texttt{Neural VC} consistently achieved near-random speaker-verification performance (EER 0.43–0.51), whereas \texttt{Resyn} showed greater variability across datasets (0.34–0.54). Both methods improved signal cleanliness (improved SNR), and \texttt{Resyn} further achieved the highest perceptual quality and lowest transcription error rates. However, these improvements were accompanied by substantial losses in pitch contour preservation(lower PCC$_\text{F0}$ values). Therefore, although anonymisation methods exhibited distinct behaviours, they consistently reflected the trade-off between preservation of speech information and effective speaker anonymisation.

\textbf{Task-dependent effects on downstream clinical classification performance.} In contrast to the largely method-dependent patterns in speaker anonymisation and the preservation of general speech characteristics, changes in downstream clinical classification performance varied substantially across tasks (Fig.~\ref{fig:metrics}g). 
Voice disorder tasks were particularly sensitive to anonymisation. Across the Saarbrücken subsets and the PUMCH dataset, \texttt{Neural VC} and \texttt{Resyn} caused significant performance degradation, with \texttt{Resyn} reducing classification accuracy to nearly random levels. \texttt{Formant$_\text{LPC}$} and \texttt{Formant$_\text{spec}$} generally produced smaller declines, suggesting that local acoustic transformations are more likely to preserve fine-grained cues related to abnormal phonation and articulatory impairment. 
Mental health and cognitive impairment tasks generally showed greater tolerance to anonymisation. Their diagnostic cues may rely more on linguistic content, pause patterns, or broader prosodic characteristics than highly detailed acoustic abnormalities, allowing multiple anonymisation methods to preserve classification performance.
The speech elicitation paradigms also affected the choice of anonymisation methods. In reading-speech datasets with fixed content, such as Androids-R and Ivanova, \texttt{Resyn} led to significant classification degradation. Resynthesising speech from transcriptions resets speaking rhythm, prosody, and abnormal voice characteristics, thereby removing informative cues which are relevant to disease recognition.

These findings highlight two considerations for voiceprint anonymisation in clinical speech. First, stronger speaker anonymisation does not necessarily lead to a better outcome, as methods that provide greater identity protection may also disrupt clinically relevant speech information, whereas methods that better preserve speech characteristics may offer weaker speaker protection. Second, the consequences of such information loss on clinical utility are task-dependent, as different disorders and elicitation paradigms rely on distinct speech characteristics. These findings motivate CASA's task-adaptive optimisation of voiceprint anonymisation strategies to achieve sufficient speaker privacy while preserving the speech information required for each clinical task.

\subsection*{Anonymisation of sensitive spoken content}

Voiceprint anonymisation alone does not address sensitive personal information explicitly disclosed through spoken content. CASA therefore performs content anonymisation by identifying privacy-sensitive information in speech transcripts and masking the corresponding spoken content. To avoid transmitting sensitive clinical information to external model services, CASA calls a locally deployed open-source LLM, Qwen2.5-32B-Instruct~\cite{yang2024qwen2technicalreport}, for sensitive-content localisation.
We analysed CASA’s content anonymisation on two in-house datasets containing sensitive spoken information: Crisis, comprising psychological hotline conversations that may disclose personal names, residential locations and telephone numbers, and PUMCH, in which a brief clinician–participant exchange involving names and ages precedes a speech repetition task. 

\begin{figure}
    \centering
    \includegraphics[width=1.0\linewidth]{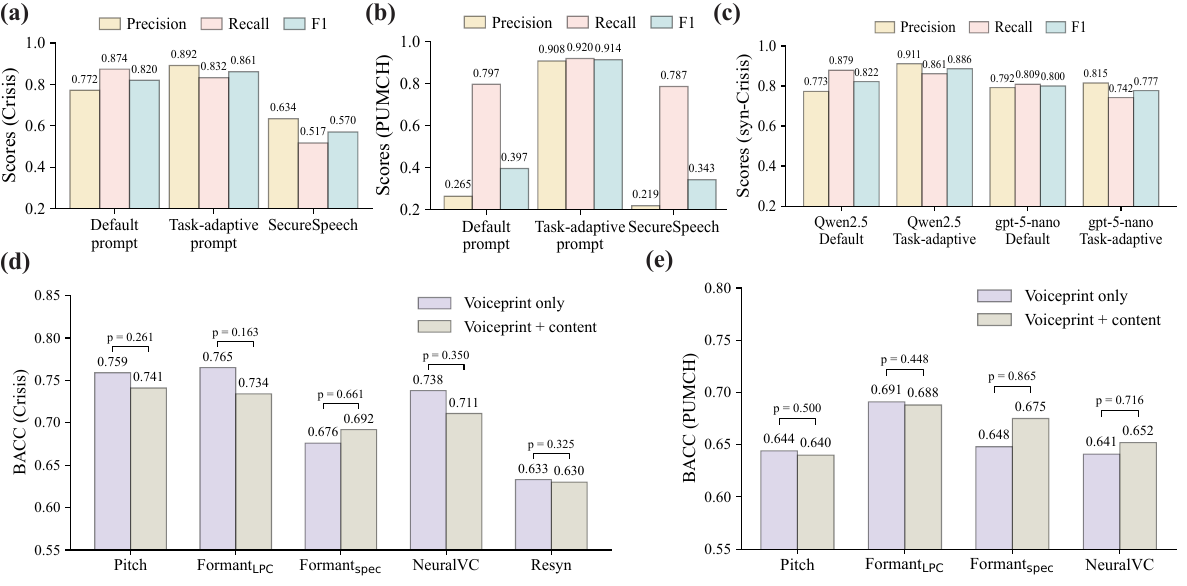}
    \caption{\textbf{$|$ Content anonymisation evaluation.} a,b, Content anonymisation on two in-house datasets, comparing CASA’s locally deployed Qwen2.5-32B-Instruct model under default and task-adaptive prompts with the NER-based localisation used in SecureSpeech. c, Comparison of Qwen2.5-32B-Instruct and GPT-5-nano on precision, recall and F1 for sensitive-content localisation. Privacy-sensitive entities were replaced with synthetic names and locations for closed-source model evaluation. d,e, Downstream task performance with and without content anonymisation. Statistical significance was assessed using one-sided exact binomial tests on discordant paired subject-level predictions, and no statistically significant reductions were observed across all evaluations. No adjustment was made for multiple comparisons.}
    \label{fig:content}
\end{figure}

We first examined CASA's performance in sensitive-content localisation. We compared CASA's LLM-based localisation with the named-entity recognition (NER)-based approach used in SecureSpeech, using manually annotated privacy-sensitive spans as ground truth. CASA outperformed SecureSpeech on both datasets (Fig.~\ref{fig:content}a,b), achieving F1 scores of 0.861 versus 0.570 on Crisis and 0.914 versus 0.343 on PUMCH.

We next investigated the contribution of task-adaptive prompting to sensitive-content localisation.
Using the same self-deployed LLM, we compared a default prompt with task-adaptive prompts refined by CASA according to dataset-specific contexts. 
The default prompt asked the model to identify common privacy-sensitive content, including names, numerical identifiers, and address information. 
The task-adaptive prompt was refined according to dataset-specific content, for example by excluding descriptions of psychological state from private identifiers in Crisis and by highlighting name-like nouns at the start of PUMCH recordings.
As shown in Fig.~\ref{fig:content}a,b, task-adaptive prompting yields marked improvement on localisation performance compared with the default prompt on both datasets with F1 scores exceeding 0.85.

We then assessed whether CASA's locally deployed open-source LLM could achieve competitive sensitive-content localisation performance compared with a closed-source model.  We compared Qwen2.5-32B-Instruct with GPT-5-nano~\cite{gpt5nano} using a synthetic-privacy version of Crisis (Fig.~\ref{fig:content}c).
To avoid exposing real private information to the closed-source model during this comparison, all privacy-sensitive spans were manually replaced with fictitious but realistic names, locations, and other identifiers, producing a synthetic-privacy  version for evaluation. Results show that Qwen2.5 achieved comparable performance as GPT-5-nano under the same synthetic-privacy setting, suggesting that the open-source Qwen2.5 LLM, when paired with appropriate prompts, is sufficient for practical content anonymisation in clinical speech.

Finally, we assessed whether content anonymisation introduced  additional losses in downstream clinical utility. We compared downstream BACC with and without content anonymisation across different voiceprint anonymisation methods on Crisis and PUMCH (Fig.~\ref{fig:content}d,e). 
No statistically significant reductions in classification performance were detected across the evaluated configurations, suggesting that privacy-sensitive spoken content can be removed without substantially affecting downstream clinical utility, enabling targeted content protection while preserving clinically relevant information.

\subsection*{Experience-guided self-improvement enables efficient task adaptation}

\begin{figure}
    \centering
    \includegraphics[width=\linewidth]{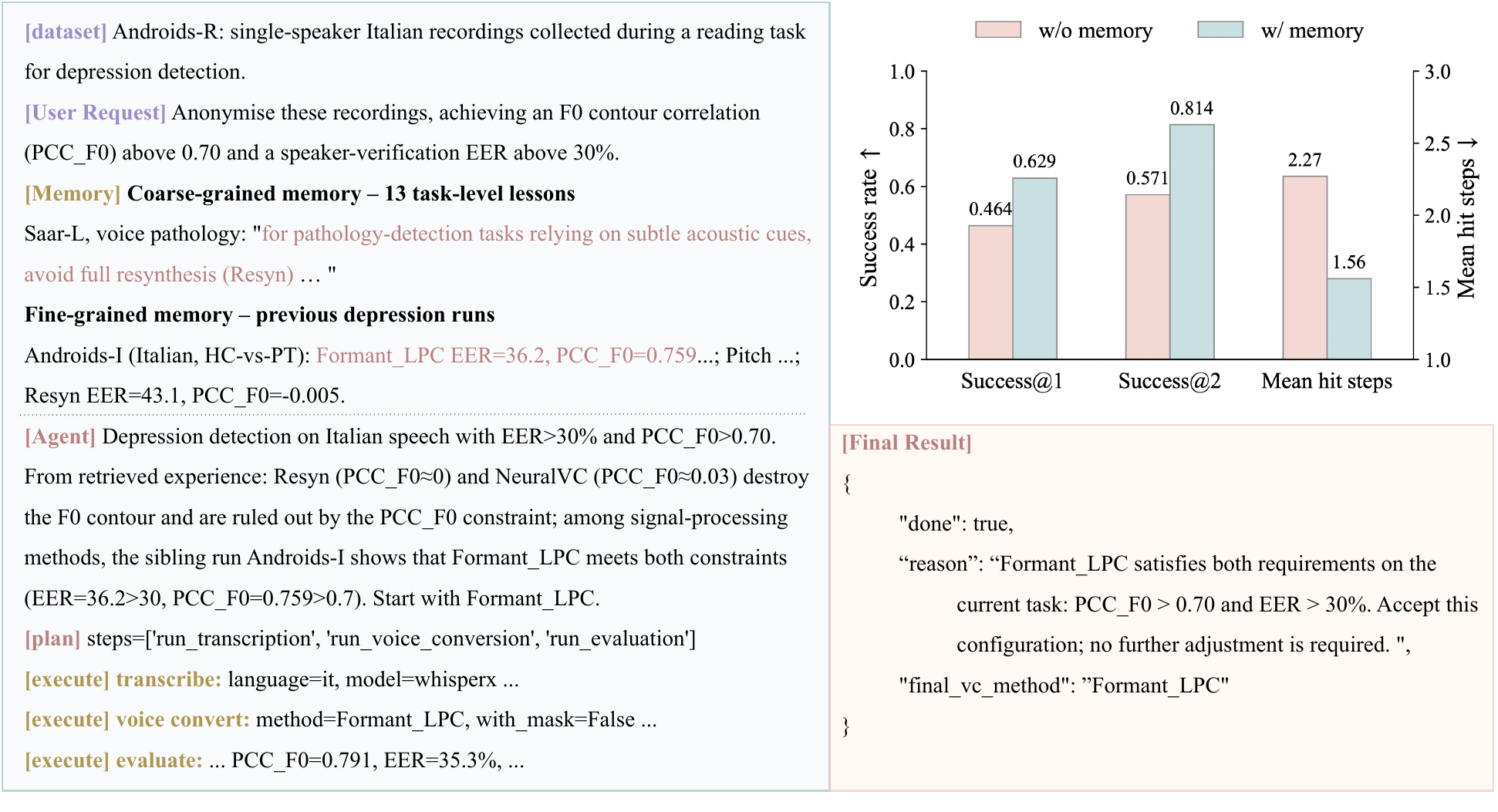}
    \caption{\textbf{$|$ Experience-guided self-improvement improves the efficiency of task adaptation.} The upper right is a comparison of CASA without and with memory. Success@1 and Success@2 (left axis) denote the proportions of runs satisfying all specified privacy and utility requirements within one and two evaluation steps, respectively. Mean hit steps (right axis) denotes the average number of steps required to first reach a configuration satisfying these requirements. Higher success rates and fewer steps indicate more efficient adaptation. The other part shows a representative execution trace for Androids-R, showing how experiences from previous runs guide CASA in subsequent run.}
    \label{fig:rsi}
\end{figure}

The preceding analyses highlight the importance of adapting anonymisation strategies to the requirements of different clinical tasks. CASA achieves this through iterative evaluation and refinement of anonymisation strategies against task-specific privacy–utility requirements. To improve the efficiency of this iterative process, CASA incorporates a recursive self-improvement mechanism that leverages experience accumulation across tasks. Specifically, resulting experience is retained in two levels of memory for subsequent planning: fine-grained memory stores anonymisation configurations and evaluation results from previous runs, and coarse-grained memory summarises task-level lessons and recurring patterns. These memories guide subsequent strategy planning, reducing the need for repeated evaluations.

To quantify the contribution of accumulated experience, we compared CASA with and without memory under the same task-specific requirements. In the memory-enabled condition, CASA could retrieve experience from other tasks, whereas the memory-free condition had no access to previous cross-task experience. Both conditions retained within-task feedback-guided adaptation. Success required satisfying all privacy and utility thresholds specified for each run. As shown in Fig.~\ref{fig:rsi}, with memory, the proportion of runs succeeding on the first evaluation increased from 46.4\% to 62.9\%, and the proportion succeeding within two evaluations increased from 57.1\% to 81.4\%. The mean number of strategy evaluations needed to reach a satisfactory configuration decreased by 31.3\%, dropping from 2.27 to 1.56 steps. These results show that accumulated cross-task experience reduces the number of anonymisation and evaluation cycles needed for task adaptation, thus enabling CASA to identify satisfactory anonymisation strategies more efficiently.

A representative example from Androids-R illustrates how accumulated experience guides strategy planning (Fig.~\ref{fig:rsi}). This involves an Italian reading task for depression detection, with user-defined requirements of PCC$_\text{F0}$ $>$ 0.70 and EER $>$ 30\%. CASA retrieved general lessons concerning the preservation of clinically relevant acoustic characteristics, together with detailed evaluation results from previous depression-related tasks. Guided by these experiences, CASA selected \texttt{Formant$_\text{LPC}$} as its first candidate. Evaluation on Androids-R yielded PCC$_\text{F0}$ = 0.791 and EER = 35.3\%, satisfying both requirements in a single trial. This example illustrates how CASA can leverage experience from related clinical tasks with different paradigms to prioritise promising anonymisation strategies, with suitability verified for the current task.

\subsection*{Clinical expert assessment of anonymised speech}

\begin{figure}
    \centering
    \includegraphics[width=1.0\linewidth]{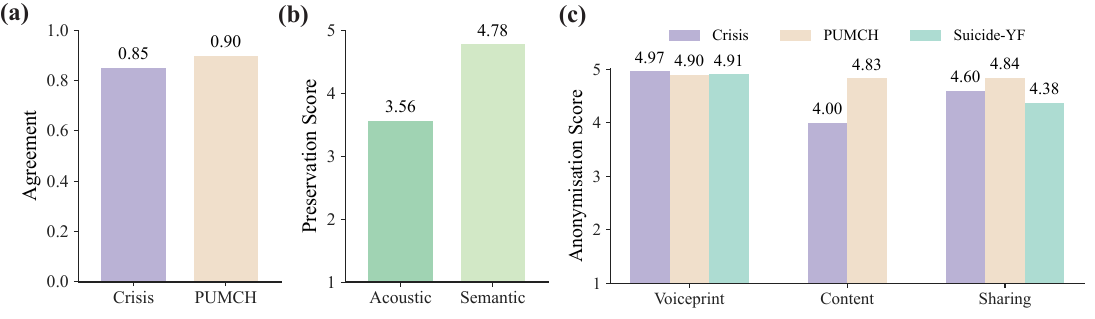}
    \caption{\textbf{$|$ Clinical expert evaluation of anonymised speech.} 
    a, Agreement between assessments of original and anonymised recordings, using crisis-level labels for Crisis and nasality-severity ratings for PUMCH.
    b, Preservation rating of acoustic and semantic information on the Suicide-YF dataset. c, Anonymisation quality and sharing acceptability rating. Clinicians rated three aspects: voiceprint anonymisation, content anonymisation, and sharing acceptability. All subjective ratings were collected on a scale of 1 to 5, with higher scores indicating better performance. Evaluators were blinded to the original reference labels.}
    \label{fig:human_eval}
\end{figure}

We further involved human experts to assess the perceived clinical usefulness of anonymised speech and its acceptability for research sharing across three clinical tasks: psychological crisis assessment, suicide risk assessment, and cleft-palate speech assessment, corresponding to the Crisis, Suicide-YF, and PUMCH datasets, respectively. 
For psychological crisis and cleft-palate speech assessment, clinicians evaluated crisis level and nasality severity, respectively. Agreement was calculated between the corresponding labels assigned from original and anonymised speech.
For suicide risk assessment, where reference labels were not derived from speech, clinicians rated the preservation of speech information relevant to suicide-risk prediction, including acoustic-feature-related and semantic-feature-related information. In addition, clinicians rated three aspects of anonymisation quality: (i) voiceprint anonymisation, which reflects whether the speaker could be tracked from voice characteristics; (ii) content anonymisation, which reflects whether sensitive spoken information was detected and masked; (iii) sharing acceptability, which reflects whether the anonymised recording was considered suitable for sharing with external researchers under a data-use agreement.

Fig.~\ref{fig:human_eval}a shows that clinical agreement remained high after anonymisation for both psychological crisis and cleft-palate speech assessment tasks. For suicide risk assessment, as shown in Fig.~\ref{fig:human_eval}b, semantic information was well preserved, with a score of 4.78, whereas acoustic information received a relatively lower score of 3.56. 
This is consistent with the voiceprint anonymisation results above, where \texttt{Resyn} preserved linguistic content but reduced preservation of original acoustic and prosodic cues.
Clinicians also gave high scores for anonymisation quality and sharing acceptability, shown in Fig.~\ref{fig:human_eval}c. Voiceprint anonymisation was rated highly across all three datasets, including PUMCH, where clinicians considered \texttt{Formant$_\text{LPC}$} to provide acceptable voice protection despite its lower EER. Content anonymisation was rated highly on PUMCH, while the score was lower on Crisis. This may be because hotline conversations are long, open-ended and context-dependent, making complete localisation of sensitive spoken information more difficult. Importantly, all three datasets received sharing acceptability scores above 4, indicating that clinicians considered the anonymised recordings suitable for controlled external research sharing. These findings provide practical evidence that the proposed anonymisation agent can help improve clinical speech data sharing while preserving information needed for downstream clinical use.

\section*{Discussion}

Clinical speech anonymisation requires balancing speaker identity protection with the preservation of information needed for clinical assessment. Voiceprint anonymisation methods that strongly suppress speaker identity may disrupt acoustic, prosodic or linguistic characteristics, whereas methods that better preserve these characteristics may offer weaker identity protection. The appropriate balance varies across clinical tasks because different disorders and speech elicitation paradigms rely on distinct speech characteristics for assessment. Our analyses showed that the same anonymisation method could preserve downstream classification performance in some tasks but substantially impair it in others. Beyond voiceprint anonymisation, clinical speech recordings may also disclose sensitive personal information through spoken content. Our content anonymisation experiments showed that identifying such information also benefits from task adaptation that accounts for differences in clinical context and speech elicitation paradigms. In sum, these findings suggest that clinical speech anonymisation needs to account for both the privacy–preservation balance and the specific requirements of different clinical tasks.

CASA addresses these challenges by reformulating clinical speech anonymisation as a task-adaptive optimisation problem, in which strategies are planned, executed and iteratively refined using multi-dimensional evaluation feedback.
Incorporating measured outcomes as feedback is necessary since the clinical relevance of individual speech characteristics cannot always be determined a priori.  Even similar clinical tasks may differ in data collection, participant populations and labelling criteria, affecting which speech characteristics are predictive of downstream outcomes. 
CASA uses downstream classification performance to directly quantify the utility of anonymised speech, reducing reliance on prior assumptions about clinically relevant speech characteristics. 
Furthermore, through recursive self-improvement, CASA accumulates successful strategies and lessons from previous evaluations to guide subsequent planning and reduce repeated exploration. This creates a continuous cycle of evaluation, experience accumulation and reuse, allowing CASA to adapt more efficiently to new clinical tasks as its experience grows.

CASA also supports human oversight of automated anonymisation. Intermediate outputs, including transcriptions, detected privacy-sensitive spans, timestamps and evaluation results, are maintained as transparent and editable representations, allowing clinical researchers and data managers to inspect, modify and approve anonymisation decisions when necessary. This human-in-the-loop design provides an additional layer of control when handling sensitive patient information. Clinical expert assessments further supported CASA's practical value, showing high agreement between clinical judgements based on original and anonymised recordings, alongside favourable ratings for anonymisation quality and research-sharing acceptability. Following CASA's anonymisation, a real-world cross-institutional clinical data sharing was conducted, where the Suicide-YF dataset was shared with Shanghai Artificial Intelligence Laboratory. By reducing speaker traceability and masking sensitive spoken content while preserving clinically relevant information, CASA could help overcome privacy-related barriers to cross-institutional clinical speech sharing. Such capabilities could facilitate collaborative research across hospitals and institutions, supporting the development of speech-based screening and assessment systems under appropriate data-use agreements.

Several limitations should be considered. First, sensitive-content localisation and voiceprint anonymisation methods such as transcription–resynthesis require automatic speech recognition (ASR) to generate speech transcripts. ASR performance may decline for low-resource languages, dialectal speech, child speech and severely pathological speech. Recognition errors can propagate to subsequent anonymisation procedures, potentially compromising privacy protection or the preservation of clinically relevant information. This limitation could be mitigated through domain-specific adaptation of ASR models for the target population or clinical setting. Second, the present study evaluated CASA using a limited set of anonymisation tools, comprising five voiceprint anonymisation methods and an LLM-based content anonymisation approach. Nevertheless, CASA's modular design allows additional anonymisation and evaluation tools to be integrated, potentially broadening its applicability to a wider range of clinical tasks. Finally, human review may remain necessary for high-risk applications where residual privacy leakage is unacceptable. To accommodate this need, CASA's human-in-the-loop design supports such oversight through transparent and editable intermediate outputs, allowing researchers and data managers to inspect, correct and approve anonymisation decisions when necessary.

Looking ahead, CASA could enable a positive feedback loop between privacy-preserving data sharing and experience-guided adaptation. By facilitating the controlled sharing of clinical speech, CASA could make more diverse data available for clinical model development, while generating anonymisation experience across a broader range of tasks. Through RSI, this accumulated experience could further improve strategy planning for new clinical tasks, potentially lowering barriers to data sharing. Future research should investigate how this reciprocal process develops across institutions and increasingly diverse clinical settings.

\section*{Methods}

We developed and evaluated CASA through a modular processing pipeline comprising transcription, voiceprint anonymisation, content anonymisation, quantitative evaluation and clinical expert assessment.

\subsection*{CASA agent}
\label{method: agent}

CASA is implemented as an LLM-based agent that plans and executes clinical speech anonymisation according to task-specific requirements. The system takes original speech recordings as input and produces anonymised recordings through voiceprint modification and sensitive-content masking. The controller receives a natural-language description of the clinical objective, speech paradigm and user-defined privacy and utility requirements. These instructions guide planning, tool execution and subsequent adaptation.

\textbf{Tools.} Transcription, voiceprint anonymisation, content anonymisation and evaluation are implemented as independent tools with explicitly defined inputs, outputs and dependency constraints (see Supplementary Fig.~\ref{fig:workflow}). Intermediate outputs are stored as structured JSON files, enabling both the agent and users to inspect and modify processing results before subsequent stages are executed.

\textbf{Memory.} The agent maintains a two-level memory to support strategy planning and feedback-guided adaptation. A coarse-grained experience memory stores summaries of previous experiments, recurring processing patterns, and failure cases with lessons learned, providing reusable prior knowledge across tasks. A fine-grained experiment memory records run-specific details, including evaluation metrics, tool parameters, and intermediate JSON outputs and logs, enabling precise diagnosis and reproduction of each anonymisation run. In addition, the current-task history is maintained and updated throughout execution, linking the two memory levels to the ongoing planning and adaptation process. After each run, the agent stores the configuration and evaluation results in fine-grained memory and distils the outcome into reusable lessons for coarse-grained memory.

\textbf{Planning and execution.} Given a natural-language anonymisation request, the agent first constructs an anonymisation strategy by selecting processing modules and configuring their corresponding parameters. Strategy generation is performed using an LLM (\textit{i.e.}, Qwen3.8-27B~\cite{qwen38}), which is provided with descriptions of available tools, parameter constraints, workflow dependencies, and relevant entries retrieved from memory. The model was run with a temperature of 0.01 and a repetition penalty of 1.05 to encourage deterministic strategy generation while reducing repetitive outputs. The resulting strategy is then executed to produce anonymised speech.

\textbf{Evaluation and adaptation.} Following execution, the evaluation module assesses the speaker privacy, downstream clinical utility and preservation of speech characteristics of the anonymised recordings, as detailed in Methods, ``Evaluation methods''. The agent subsequently analyses the evaluation result to determine whether the anonymisation objectives have been satisfied~\cite{shinn2023reflexion}. If the current strategy fails to meet the specified requirements, the agent re-plans the anonymisation strategy according to evaluation feedback, the current-task history, and previously executed processing steps retrieved from memory, and iteratively explores alternative configurations until the objectives are satisfied or a maximum of 7 adaptation iterations is reached. The agent first explores alternative anonymisation methods at their default settings and adjusts method parameters when none satisfies the specified requirements. The agent also supports manually specified anonymisation workflows if needed.

To evaluate whether the memory module improves adaptation efficiency, we constructed 10 anonymisation requests for each of the 14 datasets or task subsets, yielding 140 requests in total. Each request identified the target dataset or task subset and specified two simultaneous requirements: an EER not below 20\%, 30\% or 40\%, together with one constraint on downstream performance or speech preservation. The second constraint required a BACC decrease of less than three or six percentage points relative to original speech, a CER/WER below 15\% or 20\%, an emotion similarity above 0.93, or a pitch-contour correlation (PCC$_{\mathrm{F0}}$) above 0.70 or 0.80. We evaluated the same 140 requests with and without cross-task memory. In the memory-enabled condition, the agent could retrieve general lessons and detailed experimental records from the other 13 tasks, with experience from the task under evaluation excluded. In the memory-free condition, no cross-task experience was provided. Both conditions retained within-task planning and evaluation-guided adaptation. Success required satisfying both thresholds specified in the request. We reported Success@1 and Success@2 as the proportions of requests reaching a satisfactory configuration within one and two evaluation steps, respectively, together with the mean number of steps required to reach the first satisfactory configuration.

\subsection*{Transcription methods}
\label{method: asr}

The transcription module converts raw speech recordings into structured intermediate representations containing sentence-level transcripts, word-level timestamps, sentence indices and speaker annotations. Speech segments associated with the same speaker are aggregated and processed using a pretrained age--gender recognition model~\cite{burkhardt2023speech, age_gender_model} to infer speakers' age and gender attributes, which are subsequently stored as auxiliary metadata in the transcription files. Mandarin Chinese recordings are transcribed using Paraformer~\cite{gao2022paraformer}. Recordings in other languages are transcribed using WhisperX~\cite{bain2023whisperx}.

\subsection*{Voiceprint anonymisation methods}

The voiceprint anonymisation module supports three categories of anonymisation methods: signal-processing-based transformations, neural voice conversion, and transcription--resynthesis. Signal-processing-based methods include pitch shifting by semitone steps (\texttt{Pitch}), formant shifting through spectral scaling (\texttt{Formant$\text{spec}$}), and formant shifting through Linear Predictive Coding (LPC) pole transformation (\texttt{Formant$\text{LPC}$}) implemented by McAdams. Neural voice conversion is implemented using SeedVC, whereas transcription--resynthesis is implemented through an ASR--TTS (Text-to-Speech) cascade.

\textbf{\texttt{Pitch}.} Pitch-based anonymisation is performed using the \texttt{librosa} toolkit. According to speaker-level gender estimates obtained during transcription, a positive offset of 3 semitones is applied to male speakers and a negative offset of 3 semitones is applied to female speakers.

\textbf{\texttt{Formant$_\text{spec}$}.} Formant-based anonymisation is implemented by applying short-time Fourier transform analysis followed by linear interpolation-based spectral scaling with a scaling factor of 1.1. This operation approximates modifying the effective vocal tract length of the speaker.

\textbf{\texttt{Formant$_\text{LPC}$}.} McAdams anonymisation~\cite{patino2021speaker} leverages Linear Predictive Coding (LPC) to alter vocal tract characteristics by shifting the formant locations. For each frame, the angles of the LPC poles are transformed by a predefined power coefficient, and the speech is then resynthesised from the original residual signal with the modified LPC filter, changing the distribution of formant frequencies. An LPC order of 20 and a power coefficient of 0.8 were used.

\textbf{\texttt{Neural VC}.} Neural voice conversion anonymises speech by modifying the original waveform or acoustic features. We perform neural VC using the SeedVC~\cite{liu2024zero} framework. For multi-speaker recordings, speaker labels obtained during the transcription stage are used to assign different target speaker embeddings sampled from the AISHELL-1~\cite{bu2017aishell} corpus to different speakers within the same recording. The converted speech preserves the original temporal alignment and therefore does not alter timestamp annotations.

\textbf{\texttt{Resyn}.} Transcription--resynthesis generates anonymised speech directly from transcribed text. We used FishSpeech~\cite{liao2024fish} as the TTS backbone, with language-matched reference utterances from Mozilla Common Voice~\cite{ardila2020common} and Multilingual LibriSpeech~\cite{pratap2020mls} used to specify target speaker identities.

\subsection*{Content anonymisation methods}

The content anonymisation module involves two stages: privacy localisation and audio redaction. The privacy localisation stage identifies privacy-sensitive linguistic spans from transcription output and generates structured annotations containing detected text segments together with their corresponding timestamp ranges. Localisation is implemented using an LLM. By default, a locally deployed Qwen2.5-32B-Instruct model is used to avoid transmitting transcripts to external services and privacy leakage. The model is prompted to identify common forms of spoken private information, including names, numerical identifiers, and address information. To support long-form recordings, transcription outputs are segmented into groups of consecutive sentences before localisation.

The localisation module supports configurable prompts and model backends. Task-specific prompts can be provided to adapt privacy localisation to different recording protocols, clinical settings and languages.
Alternative locally deployed LLMs can also be integrated to support different deployment environments.

Following privacy localisation, audio redaction is performed according to the anonymisation method. For methods that preserve timestamp alignment (\texttt{Pitch}, \texttt{Formant$_\text{spec}$}, \texttt{Formant$_\text{LPC}$}, and \texttt{Neural VC}), detected privacy-sensitive spans are redacted in the voiceprint-anonymised audio by replacing the corresponding speech segments with noise. For \texttt{Resyn}, privacy localisation and masking are first applied to the transcription, after which anonymised speech is generated from the modified text.

\subsection*{Evaluation methods}
\label{sec: method-eval}

We build a multi-dimensional evaluation framework to comprehensively assess the effectiveness of anonymisation methods.

\textbf{Speech quality.} Speech quality is evaluated using signal-to-noise ratio (SNR) and mean opinion score (MOS). SNR measures cleanliness of the speech while MOS measures the overall perceptual quality of the speech. MOS is automatically estimated using UTMOS~\cite{saeki2022utmos}.

\textbf{Contour preservation}. Pitch preservation is evaluated through fundamental frequency (F0) contour analysis. F0 contours are extracted from both original and anonymised speech using the \texttt{librosa} toolkit.
Preservation of pitch-contour shape is quantified using the Pearson correlation coefficient (PCC) between the two contours. 

\textbf{Linguistic preservation.} Linguistic preservation is evaluated by comparing transcriptions generated from the original and anonymised speech. For Mandarin Chinese recordings, preservation is quantified using character error rate (CER). For other languages, word error rate (WER) is used.

\textbf{Emotion preservation.} Emotion preservation is evaluated using Emotion2Vec-Plus-Seed~\cite{ma2024emotion2vec}. Emotion embeddings are extracted from both original and anonymised speech, and preservation is quantified using cosine similarity between the corresponding embeddings.

\textbf{Speaker anonymisation.} Speaker anonymisation is evaluated using equal error rate (EER), with higher EER indicating weaker speaker traceability. Speaker embeddings are extracted using the WavLM-Base-Plus-sv speaker verification model~\cite{wavlm-sv}. Positive verification pairs consist of original and anonymised recordings from the same subject, whereas negative pairs consist of original and anonymised recordings from different subjects. Verification scores are computed as cosine similarities between speaker embeddings. 
EER was estimated at the ROC operating point that minimised the absolute difference between the false acceptance and false rejection rates, and was calculated as the mean of these two rates at that point. The 95\% confidence interval was estimated using a nonparametric percentile bootstrap with 1,000 resampling iterations, with EER recalculated for each resample, and the 2.5th and 97.5th percentiles of the valid bootstrap estimates defined the confidence interval.
The framework supports user-provided subject mappings, allowing multiple recordings to be treated as originating from the same subject during verification. When no subject mapping is provided, each recording is treated as an independent subject.

\textbf{Clinical utility.} Downstream clinical utility is evaluated through disease-related classification. We use the speech LLM Qwen2.5-Omni-7B~\cite{xu2025qwen2} as the backbone of an end-to-end classification model, following our previous work~\cite{cui2026speaker} (Supplementary Fig.~\ref{fig:downstream_model}). The model consists of a speech encoder, an LLM, and a fully connected classification head that maps the final representation to disease-label predictions. Weight-decomposed low-rank adaptation (DoRA)~\cite{liu2024dora} is used for parameter-efficient fine-tuning with a rank of 32 and an alpha of 64. The learning rate is set to $1 \times 10^{-4}$, and the number of training epochs is adjusted according to dataset size. To reduce the influence of class imbalance, minority classes in the training set are upsampled until all classes contain the same number of samples. Balanced accuracy (BACC) is used as the primary metric for downstream utility.
Long recordings are divided into non-overlapping 4-minute segments, with each treated as a separate sample during training and inference. The final recording-level prediction is obtained by averaging segment-level class probabilities at test time.
For each dataset, we trained a separate classifier on the original speech and on each anonymised version, using the corresponding speech version for both training and evaluation.

For datasets with official test splits, evaluation is performed on the official test set using five independent runs with different random seeds. For datasets without official splits, five-fold cross-validation is used. In both settings, train, validation, and test sets are strictly subject-independent to avoid data leakage and to evaluate cross-subject generalisation. Mean BACC and standard deviation are reported across the five runs or folds. 
Statistical significance of utility degradation after anonymisation was assessed using one-sided exact binomial tests on discordant paired predictions, using \texttt{scipy.stats.binomtest}.
No adjustment was made for multiple comparisons. A lower $P$ value indicates stronger evidence that anonymisation reduces downstream classification performance. Statistical significance was defined as $P<0.05$.

\subsection*{Datasets for validation}
\label{sec: method dataset}

\textbf{MODMA}~\cite{cai2022multi} is a Mandarin Chinese clinical speech dataset comprising 1,503 recordings from 52 participants aged 16--56 years, including outpatients diagnosed with depression (44.2\%) and healthy controls (55.8\%). Speech was collected through multiple elicitation paradigms, including interview, reading and picture-description tasks.

\textbf{Androids-Corpus}~\cite{tao2023androids} comprises recordings from 118 native Italian speakers classified as healthy controls (HC) or patients (PT) with depression-related disorders. Reading recordings (54 HC, 58 PT) were collected using Aesop’s fable \textit{The North Wind and the Sun}, whereas interview recordings (52 HC, 64 PT) were obtained through spontaneous responses to daily-life questions.

\textbf{Suicide-YF} consists of Mandarin Chinese speech recordings from 1,223 adolescents aged 10--18 years recruited in Guangdong, China. 
The participants were asked to complete an online survey, in which they were asked whether they had ever experienced suicidal thoughts or behaviours. Those who answered yes were invited to participate in the study. Additionally, a control group of children with no suicide risk was selected for comparison with the at-risk participants.
All recordings are paired with suicide-risk assessments administered using the Mini International Neuropsychiatric Interview for Children and Adolescents (MINI-KID) suicidality module~\cite{sheehan2010reliability}, with 53.4\% of participants identified as at risk. Following previous studies~\cite{cui2024spontaneous, cui2026speaker}, only recordings from the self-description task were used. Informed consent was obtained from all participants and their guardians, and the study was approved by the Tsinghua University Institutional Review Board.

\textbf{Crisis} is an in-house Mandarin Chinese psychological-support hotline dataset comprising 154 recordings (approximately 100 hours in total; mean duration 39.49 minutes). 
The study was conducted at the Beijing Psychological Support Hotline which provides services to callers from around China, three-fourths of whom live outside of the Beijing Municipality. Hotline operators are full-time employees with a psychology or psychiatry background who receive at least 3 months full-time training in the provision of hotline services before independently taking calls.
Recordings were independently annotated by two clinical experts into three crisis levels: no crisis (55 participants), low crisis (42), and medium-to-high crisis (57). Conversations cover a broad range of personal issues, including depression, anxiety, relationship stress and suicidal ideation, and may contain privacy-sensitive information such as names, locations and phone numbers.
The study was approved by the Institutional Review Board of Beijing Huilongguan Hospital (2014-106, 2024-76). Before callers were connected to a hotline operator, they were informed by a voice message that all calls of the hotline would be tape-recorded and that data utilised for research purposes would be analysed anonymously. 

\textbf{NCMMSC-AD}~\cite{chen2023raw} is a Mandarin Chinese cognitive assessment dataset comprising 399 recordings from 175 participants labelled as healthy controls (HC), mild cognitive impairment (MCI), or Alzheimer's disease (AD). Recordings include multiple cognitive tasks, including verbal fluency, lexical retrieval, picture description and self-introduction.

\textbf{ADReSS}~\cite{luz2021alzheimer} is an English Alzheimer's disease dataset derived from the Pitt corpus, comprising recordings from 156 participants aged 50--79 years. The official split contains 54 HC and 54 AD participants for training, and 24 HC and 24 AD participants for testing. All recordings correspond to the Cookie Theft picture-description task.

\textbf{Ivanova}~\cite{ivanova2022discriminating} is a Spanish cognitive assessment dataset comprising recordings from 357 participants aged 50--96 years. Participants were labelled as healthy controls (54.9\%), mild cognitive impairment (24.4\%), or Alzheimer's disease (20.7\%). All recordings correspond to a reading task in which participants read the first two sentences of \textit{Don Quixote} by Cervantes.

\textbf{NeuroVoz}~\cite{mendes2024neurovoz, neurovoz_zenodo} is a Spanish Parkinson's disease dataset comprising recordings from 112 participants aged 31--88 years, including 58 healthy controls and 54 patients with Parkinson's disease. The corpus includes multiple speech tasks, including sustained vowels, diadochokinetic tests, listen-and-repeat utterances, and spontaneous monologues.

\textbf{ENNI}~\cite{schneider2006storytelling} is an English narrative-speech dataset from the CHILDES Clinical English Edmonton Narrative Norms Instrument corpus, comprising recordings from Canadian children with typical development (286) or specific language impairment (66). The task is picture-elicited story generation, in which participants narrate a picture story to an examiner.

\textbf{Saarbrücken}~\cite{putzer2008saarbruencken} is a German pathological-voice dataset containing clinically validated voice-disorder labels. 
We selected three common voice-disorder categories from the database and constructed three disease-versus-control subsets from the read-sentence recordings: laryngitis (140 healthy, 140 pathological), Rekurrensparese (212 healthy, 212 pathological), and hyperfunctional dysphonia (212 healthy, 212 pathological). 
All recordings correspond to the sentence \textit{``Guten Morgen, wie geht es Ihnen?'' (``Good morning, how are you?'')}

\textbf{PUMCH} is an in-house Mandarin Chinese speech dataset for cleft-palate speech assessment, comprising recordings from 244 participants, including 128 patients with cleft palate and 116 healthy controls.
Synchronized audio-video recordings were obtained primarily using an iPhone 16 in a quiet hospital room; additional home recordings were submitted using consumer devices under variable recording conditions. A researcher was present during hospital-based collection, and a parent or guardian could accompany minors when necessary; no researcher was present during home recording.
Disease labels were annotated by experienced clinicians. Recordings consist of clinician–participant interactions beginning with identity-related questions (\textit{e.g.}, name and age), followed by a sentence-repetition task in which participants repeat utterances read by the clinician. This multicentre, prospective observational study was approved by the Ethics Committee of Peking Union Medical College Hospital, Chinese Academy of Medical Sciences, Beijing, China (approval number: I-25PJ0105). 
For adult participants, written informed consent was obtained from them directly; for minors, written informed consent was obtained from their parents or legal guardians, in accordance with the Declaration of Helsinki.

\subsection*{Fixed pipeline for baseline}

To evaluate whether adaptive anonymisation provides advantages over conventional fixed strategies, we implemented SecureSpeech as a representative fixed anonymisation baseline. 
SecureSpeech adopts a sequential ASR--NER--LLM--TTS pipeline, where ASR is first applied to obtain transcripts, NER identifies privacy-sensitive entities, an LLM rewrites the detected spans, and a TTS model generates anonymised speech.

Since the original SecureSpeech implementation was developed for English speech only and its individual components do not directly support multilingual clinical speech evaluation, we adapted the pipeline for our experimental setting. 
Specifically, to enable privacy entity detection across multiple languages, the original English-only NER module was replaced with a RoBERTa-based NER model fine-tuned on the CLUENER2020 dataset~\cite{roberta_zh_ner} for Chinese speech processing, and a multilingual BERT-based NER model~\cite{Davlan2026bert_ner} trained on high-resource languages for multilingual scenarios.
The original ParlerTTS-based synthesis module was replaced with FishSpeech to provide multilingual speech resynthesis capability. 
The LLM component for rewriting was replaced with locally deployed Qwen2.5-32B-Instruct. 
The ASR outputs were kept identical to those used in our framework to ensure a fair comparison.

\subsection*{External attacker evaluation}

We evaluated CASA’s anonymised recordings using two external attackers: a semi-informed attacker trained on anonymised speech following the VoicePrivacy training protocol, and a SpeechLLM attacker using a pretrained speech–language model directly for speaker verification. Both attackers compared original and anonymised recordings to assess whether speaker identities remained linkable. Their outputs were used for external evaluation and were not provided to CASA during planning or adaptation.

\textbf{Semi-informed attacker.} We trained an ECAPA-TDNN~\cite{desplanques2020ecapa} speaker-verification model on an anonymised version of LibriSpeech train-clean-360~\cite{panayotov2015librispeech}. Because the intended use of LibriSpeech is automatic speech recognition, CASA was instructed to preserve linguistic information with a CER/WER requirement below 10\%. The anonymised training recordings were generated using CASA’s \texttt{Resyn} method. Model training followed the VoicePrivacy 2024 training recipe~\cite{tomashenko2026third}. The resulting attacker was then applied to the anonymisation configurations produced by CASA for the clinical tasks. EER was calculated using the same pairing and scoring procedure described in ``Evaluation methods'', replacing the agent’s speaker-embedding model with the ECAPA-TDNN attacker trained on CASA-anonymised LibriSpeech.

\textbf{SpeechLLM attacker.} We used SALMONN-2-8B~\cite{yang2026salmonn} directly to classify pairs of recordings as originating from the same or different speakers. For each speaker shared between the original and anonymised recordings, we generated one same-speaker pair and one different-speaker pair, yielding 2N pairs for N speakers. For every pair, the model received the first 15 seconds of the original recording and the last 15 seconds of the anonymised recording. When an individual recording was shorter than 30 seconds, its first or second half was used, respectively. A fixed prompt asked whether the main speakers in the two recordings were the same person and required exactly one response: “SAME” or “DIFFERENT”. Each pair was processed independently, with sampling disabled and a maximum output length of 256 tokens. The verification error, calculated as $1 - \text{accuracy}$ was reported. 

\subsection*{Clinical expert evaluation}

We conducted clinical expert evaluations on three private clinical speech datasets to assess whether anonymised recordings preserved clinically relevant information and were acceptable for controlled research sharing. The expert panel comprised clinicians with relevant clinical experience and researchers extensively trained in the corresponding clinical domains. Evaluation protocols were tailored to the characteristics and assessment objectives of each clinical task. Throughout the clinical expert evaluations, evaluators were blinded to the original reference labels.

For the crisis assessment task, we randomly selected 60 recordings from the Crisis dataset for clinical relabelling, as the relatively long duration of the recordings made exhaustive expert evaluation impractical. Only the anonymised versions were presented to the experts, who independently assigned crisis labels following the same assessment procedure used to construct the original dataset. Labels derived from anonymised speech were compared with the original reference labels, and agreement was quantified as the proportion of matching labels. We additionally sampled a non-overlapping set of 30 recordings to assess anonymisation quality and acceptability for research sharing. For this evaluation, experts were provided with paired original and anonymised recordings. Two clinical experts jointly participated in the evaluation; each relabelled 30 anonymised recordings and assessed 15 original–anonymised pairs for anonymisation quality and sharing acceptability. The web interface is shown in Supplementary Fig.~\ref{fig:web_crisis_util} and Supplementary Fig.~\ref{fig:web_crisis_anony}.

For the suicide risk assessment task, 350 recordings were randomly selected from the Suicide-YF dataset and evaluated by one expert. Because the reference suicide-risk labels were derived from clinical assessments independent of the speech recordings, the expert was not asked to relabel the anonymised speech. Instead, paired original and anonymised recordings were presented to assess whether clinically relevant information for suicide-risk prediction was preserved after anonymisation. Preservation was evaluated along two dimensions: acoustic-feature-related information and semantic-feature-related information. As content anonymisation was not applied in this task, the anonymisation-quality assessment focused on voiceprint anonymisation and acceptability for research sharing. The web interface is shown in Supplementary Fig.~\ref{fig:web_suicide}.

For the cleft-palate speech assessment task, all 244 recordings from the PUMCH dataset were evaluated by a speech-language pathologist. To simplify the evaluation procedure, rather than performing a comprehensive cleft-palate speech assessment covering multiple articulation and resonance errors, the evaluation focused on nasality severity as a clinically relevant perceptual feature, rated on a four-level ordinal scale: none, mild, moderate, and severe. Clinical information preservation was quantified by the exact agreement between ratings obtained from the original and anonymised recordings. The speech-language pathologist also assessed voiceprint anonymisation, content anonymisation, and acceptability for research sharing for each anonymised recording. The web interface is shown in Supplementary Fig.~\ref{fig:web_pumch}.

Preservation, anonymisation-quality and sharing-acceptability scores were collected on a five-point scale, with higher scores indicating better preservation, stronger anonymisation or higher sharing acceptability. Mean scores were reported across evaluated recordings.

\subsection*{Reporting summary}

Further information on research design is available in the Nature Research
Reporting Summary linked to this article.

\section*{Data availability}

All public datasets used in this study are available from their original sources, which are cited in the manuscript. The three in-house datasets, namely Suicide-YF, Crisis and PUMCH, are not publicly available because they contain sensitive clinical speech recordings. These datasets may be made available upon reasonable request to the corresponding authors, subject to the respective data-use agreements and ethical approval requirements.

\section*{Code availability}

The code is publicly available at \url{https://github.com/JaderOx/AnonymiseAgent} under the Apache-2.0 license.

\bibliography{sn-bibliography}

\begin{appendices}

\section*{Acknowledgements}

This work was supported by the National Natural Science Foundation of China (Grant No. 62501336) and CAMS Innovation Fund for Medical Sciences (Grant No. 2023-I2M-C\&T-B-050).

\section*{Author contributions}

Z.C., W.W., and C.Z. designed the research; Z.C., H.N., H.S., and S.Z. performed the research; Y.L., R.C., L.Z., and Y.T. provided datasets; Z.C., H.N., W.W., and C.Z. analysed the data; J.W., L.C., C.L., Y.Z., C.S., and Y.T. contributed to human evaluation; and Z.C., H.N., W.W., and C.Z. wrote the paper. All authors read and approved the final manuscript.

\section*{Competing interests}

The authors declare no competing interests.

\newpage
\section*{Supplementary Information}
\label{sec:supplementary_information}

\setcounter{figure}{0}
\setcounter{table}{0}
\renewcommand{\thefigure}{\arabic{figure}}
\renewcommand{\thetable}{\arabic{table}}
\renewcommand{\figurename}{Supplementary Fig.}
\renewcommand{\tablename}{Supplementary Table}

\begin{sidewaystableplain}[h]
\caption{\textbf{$|$ Results on mental health disorder datasets.} EER is reported with its 95\% CI. Other metrics except CER/WER are reported as mean $\pm$ standard deviation. For BACC, p-values were computed using one-sided exact binomial tests on discordant paired subject-level predictions, comparing the predictions from anonymised and original speech. No adjustment was made for multiple comparisons. Results with $p < 0.05$ are shown in bold.}
\label{tab: result1}
\resizebox{1\linewidth}{!}{
\begin{tabular}{@{}cccccccccc@{}}
\toprule
Dataset                           & Anonymise                & SNR                               & MOS                              & PCC$_\text{F0}$                    & CER/WER                & Emo                                & EER                                    & BACC                            & p-value          \\ \midrule
\multirow{6}{*}{MODMA}            & Origin                   & 20.58 $\pm$ 5.36                  & 2.25 $\pm$ 0.50                  & -                                  & -                      & -                                  & -                                      & 0.782 $\pm$ 0.078           & -                \\
                                  & Pitch                    & 19.01 $\pm$ 5.28                  & 1.27 $\pm$ 0.03                  & 0.903 $\pm$ 0.148                  & 0.204                  & 0.934 $\pm$ 0.068                  & 0.567 (0.529 - 0.619)                    & 0.787 $\pm$ 0.126           & 0.746            \\
                                  & Formant$_\text{LPC}$                  & 22.39 $\pm$ 5.51                  & 1.54 $\pm$ 0.31                  & 0.851 $\pm$ 0.300                  & 0.117                  & 0.944 $\pm$ 0.063                  & 0.139 (0.125 - 0.153)                    & 0.807 $\pm$ 0.073           & 0.746            \\
                                  & Formant$_\text{spec}$                  & 19.00 $\pm$ 5.17                  & 1.31 $\pm$ 0.07                  & 0.953 $\pm$ 0.097                  & 0.098                  & 0.952 $\pm$ 0.056                  & 0.163 (0.149 - 0.177)                    & 0.795 $\pm$ 0.146           & 0.604            \\
                                  & Neural VC                   & 36.85 $\pm$ 5.36                  & 2.00 $\pm$ 0.37                  & 0.279 $\pm$ 0.438                  & 0.170                  & 0.936 $\pm$ 0.066                  & 0.445 (0.426 - 0.465)                    & 0.773 $\pm$ 0.099           & 0.604            \\
                                  & Resyn                  & 30.78 $\pm$ 8.29                  & 3.23 $\pm$ 0.36                  & 0.070 $\pm$ 0.272                  & 0.040                  & 0.928 $\pm$ 0.074                  & 0.335 (0.318 - 0.351)                    & 0.842 $\pm$ 0.103           & 0.849            \\ \midrule
\multirow{6}{*}{Androids-I}
    & Origin                & 22.72 $\pm$ 4.54 & 1.52 $\pm$ 0.19 & -                      & -     & -                     & -                   & 0.853 $\pm$ 0.090          & -              \\
    & Pitch                 & 22.04 $\pm$ 4.37 & 1.25 $\pm$ 0.01 & 0.441 $\pm$ 0.355      & 0.189 & 0.962 $\pm$ 0.051     & 0.319 (0.259 - 0.388) & 0.852 $\pm$ 0.092          & 0.696          \\
    & Formant$_\text{LPC}$   & 21.16 $\pm$ 4.54 & 1.26 $\pm$ 0.04 & 0.759 $\pm$ 0.233      & 0.187 & 0.914 $\pm$ 0.074     & 0.362 (0.306 - 0.430) & 0.800 $\pm$ 0.093          & 0.090          \\
    & Formant$_\text{spec}$  & 21.26 $\pm$ 4.48 & 1.26 $\pm$ 0.05 & 0.412 $\pm$ 0.362      & 0.125 & 0.950 $\pm$ 0.046     & 0.177 (0.134 - 0.238) & 0.825 $\pm$ 0.075          & 0.274          \\
    & Neural VC             & 40.73 $\pm$ 1.87 & 1.72 $\pm$ 0.11 & 0.033 $\pm$ 0.149      & 0.258 & 0.935 $\pm$ 0.059     & 0.483 (0.410 - 0.550) & \textbf{0.774 $\pm$ 0.088} & \textbf{0.047} \\
    & Resyn                 & 39.02 $\pm$ 4.33 & 2.44 $\pm$ 0.46 & -0.005 $\pm$ 0.102     & 0.021 & 0.957 $\pm$ 0.061     & 0.431 (0.367 - 0.500) & 0.896 $\pm$ 0.051          & 0.593          \\ \midrule
\multirow{6}{*}{Androids-R}
    & Origin                & 22.89 $\pm$ 5.04 & 1.71 $\pm$ 0.29 & -                      & -     & -                     & -                   & 0.699 $\pm$ 0.088          & -              \\
    & Pitch                 & 21.69 $\pm$ 4.89 & 1.25 $\pm$ 0.02 & 0.878 $\pm$ 0.183      & 0.079 & 0.957 $\pm$ 0.058     & 0.330 (0.250 - 0.380) & 0.640 $\pm$ 0.084          & 0.229          \\
    & Formant$_\text{LPC}$   & 21.39 $\pm$ 4.84 & 1.29 $\pm$ 0.05 & 0.791 $\pm$ 0.188      & 0.067 & 0.891 $\pm$ 0.066     & 0.353 (0.295 - 0.424) & 0.675 $\pm$ 0.082          & 0.500          \\
    & Formant$_\text{spec}$  & 22.12 $\pm$ 5.12 & 1.27 $\pm$ 0.05 & 0.808 $\pm$ 0.250      & 0.036 & 0.962 $\pm$ 0.053     & 0.232 (0.183 - 0.295) & 0.745 $\pm$ 0.110          & 0.952          \\
    & Neural VC             & 39.82 $\pm$ 2.50 & 1.81 $\pm$ 0.14 & 0.052 $\pm$ 0.142      & 0.084 & 0.948 $\pm$ 0.066     & 0.509 (0.433 - 0.572) & 0.724 $\pm$ 0.061          & 0.837          \\
    & Resyn                 & 35.50 $\pm$ 4.40 & 2.24 $\pm$ 0.39 & 0.017 $\pm$ 0.114      & 0.010 & 0.940 $\pm$ 0.069     & 0.460 (0.384 - 0.531) & \textbf{0.516 $\pm$ 0.063} & \textbf{0.008} \\ \midrule
\multirow{6}{*}{Suicide-YF}       & Origin                   & 15.22 $\pm$ 4.30                  & 1.44 $\pm$ 0.25                  & -                                  & -                      & -                                  & -                                      & 0.668 $\pm$ 0.010           & -                \\
                                  & Pitch                    & 14.21 $\pm$ 4.04                  & 1.31 $\pm$ 0.09                  & 0.702 $\pm$ 0.290                  & 0.365                  & 0.922 $\pm$ 0.074                  & 0.523 (0.502 - 0.541)                     & \textbf{0.604 $\pm$ 0.025}  & $\mathbf{1.59\times 10^{-3}}$ \\
                                  & Formant$_\text{LPC}$                  & 18.62 $\pm$ 4.62                  & 1.28 $\pm$ 0.08                  & 0.545 $\pm$ 0.448                  & 0.208                  & 0.946 $\pm$ 0.060                  & 0.180 (0.167 - 0.197)                     & 0.641 $\pm$ 0.013           & 0.069            \\
                                  & Formant$_\text{spec}$                  & 14.33 $\pm$ 4.13                  & 1.31 $\pm$ 0.12                  & 0.844 $\pm$ 0.228                  & 0.149                  & 0.953 $\pm$ 0.055                  & 0.230 (0.212 - 0.246)                     & 0.645 $\pm$ 0.017           & 0.074            \\
                                  & Neural VC                   & 33.52 $\pm$ 6.04                  & 1.82 $\pm$ 0.30                  & 0.126 $\pm$ 0.330                  & 0.336                  & 0.922 $\pm$ 0.073                  & 0.467 (0.449 - 0.489)                     & \textbf{0.629 $\pm$ 0.023}  & \textbf{0.014}   \\
                                  & Resyn                  & 33.17 $\pm$ 8.29                  & 2.93 $\pm$ 0.40                  & 0.013 $\pm$ 0.218                  & 0.020                  & 0.915 $\pm$ 0.076                  & 0.412 (0.393 - 0.431)                     & 0.663 $\pm$ 0.018           & 0.376            \\ \midrule
\multirow{6}{*}{Crisis}           & Origin                   & 13.47 $\pm$ 4.36                  & 1.33 $\pm$ 0.05                  & -                                  & -                      & -                                  & -                                      & 0.765 $\pm$ 0.070           & -                \\
                                  & Pitch                    & 10.73 $\pm$ 3.92                  & 1.23 $\pm$ 0.01                  & 0.809 $\pm$ 0.167                  & 0.324                  & 0.954 $\pm$ 0.054                  & 0.585 (0.549 - 0.615)                  & 0.759 $\pm$ 0.067           & 0.721            \\
                                  & Formant$_\text{LPC}$                  & 13.69 $\pm$ 4.36                  & 1.23 $\pm$ 0.01                  & 0.590 $\pm$ 0.297                  & 0.136                  & 0.951 $\pm$ 0.060                  & 0.220 (0.193 - 0.250)                  & 0.765 $\pm$ 0.066           & 0.639            \\
                                  & Formant$_\text{spec}$                  & 11.83 $\pm$ 4.11                  & 1.22 $\pm$ 0.00                  & 0.816 $\pm$ 0.192                  & 0.116                  & 0.965 $\pm$ 0.049                  & 0.187 (0.158 - 0.222)                  & \textbf{0.676 $\pm$ 0.061}  & \textbf{0.044}   \\
                                  & Neural VC                   & 40.44 $\pm$ 1.65                  & 1.64 $\pm$ 0.07                  & 0.092 $\pm$ 0.218                  & 0.197                  & 0.955 $\pm$ 0.058                  & 0.495 (0.460 - 0.534)                  & 0.738 $\pm$ 0.035           & 0.229            \\
                                  & Resyn                  & 52.74 $\pm$ 6.28                  & 2.22 $\pm$ 0.08                  & -0.007 $\pm$ 0.202                  & 0.031                  & 0.852 $\pm$ 0.036                  & 0.515 (0.481 - 0.549)                  & \textbf{0.633 $\pm$ 0.145}  & \textbf{0.032}   \\ \bottomrule

\end{tabular}
}
\end{sidewaystableplain}

\begin{sidewaystableplain}[]
\centering
\caption{\textbf{$|$ Results on cognitive disorder datasets.}}
\label{tab: result2}
\resizebox{1\linewidth}{!}{
\begin{tabular}{@{}cccccccccc@{}}
\toprule
Dataset                    & Anonymise & SNR               & MOS             & PCC$_\text{F0}$    & CER/WER & Emo               & EER                   & BACC                           & p-value          \\ \midrule
\multirow{6}{*}{NCMMSC-AD} & Origin    & 24.38 $\pm$ 6.54  & 1.38 $\pm$ 0.14 & -                  & -       & -                 & -                     & 0.765 $\pm$ 0.024             & -                \\
                           & Pitch     & 22.87 $\pm$ 6.57  & 1.26 $\pm$ 0.03 & 0.783 $\pm$ 0.192  & 0.404   & 0.931 $\pm$ 0.073 & 0.576 (0.539 - 0.615) & \textbf{0.688 $\pm$ 0.024}             & \textbf{0.016}            \\
                           & Formant$_\text{LPC}$   & 26.45 $\pm$ 6.59  & 1.25 $\pm$ 0.03 & 0.485 $\pm$ 0.476  & 0.273   & 0.941 $\pm$ 0.062 & 0.147 (0.124 - 0.175) & 0.730 $\pm$ 0.029             & 0.058       \\
                           & Formant$_\text{spec}$   & 23.17 $\pm$ 6.49  & 1.24 $\pm$ 0.02 & 0.832 $\pm$ 0.215  & 0.221   & 0.960 $\pm$ 0.052 & 0.190 (0.165 - 0.221) & 0.799 $\pm$ 0.014             & 0.639            \\
                           & Neural VC    & 39.19 $\pm$ 2.77  & 1.73 $\pm$ 0.17 & 0.128 $\pm$ 0.268  & 0.365   & 0.942 $\pm$ 0.069 & 0.467 (0.432 - 0.502) & \textbf{0.681 $\pm$ 0.007}             & \textbf{0.002}            \\
                           & Resyn   & 30.07 $\pm$ 6.90  & 2.58 $\pm$ 0.22 & 0.012 $\pm$ 0.179  & 0.048   & 0.912 $\pm$ 0.076 & 0.479 (0.449 - 0.514) & 0.754 $\pm$ 0.013    & 0.243   \\ \midrule
\multirow{6}{*}{ADReSS}    & Origin    & 46.11 $\pm$ 9.52  & 1.58 $\pm$ 0.36 & -                  & -       & -                 & -                     & 0.757 $\pm$ 0.035             & -                \\
                           & Pitch     & 44.78 $\pm$ 9.17  & 1.27 $\pm$ 0.03 & 0.846 $\pm$ 0.165  & 0.235   & 0.963 $\pm$ 0.047 & 0.312 (0.260 - 0.368) & 0.815 $\pm$ 0.019             & 0.990            \\
                           & Formant$_\text{LPC}$   & 43.48 $\pm$ 8.91  & 1.26 $\pm$ 0.04 & 0.739 $\pm$ 0.294  & 0.354   & 0.938 $\pm$ 0.059 & 0.361 (0.304 - 0.413) & 0.711 $\pm$ 0.051             & 0.111            \\
                           & Formant$_\text{spec}$   & 43.97 $\pm$ 9.15  & 1.26 $\pm$ 0.04 & 0.859 $\pm$ 0.210  & 0.177   & 0.974 $\pm$ 0.036 & 0.148 (0.107 - 0.194) & 0.774 $\pm$ 0.018             & 0.797            \\
                           & Neural VC    & 41.00 $\pm$ 4.30  & 1.88 $\pm$ 0.37 & 0.297 $\pm$ 0.269  & 0.353   & 0.913 $\pm$ 0.050 & 0.513 (0.445 - 0.561) & 0.811 $\pm$ 0.020             & 0.985            \\
                           & Resyn   & 65.07 $\pm$ 8.55  & 3.83 $\pm$ 0.11 & 0.045 $\pm$ 0.202  & 0.025   & 0.908 $\pm$ 0.067 & 0.536 (0.441 - 0.613) & 0.754 $\pm$ 0.027             & 0.561            \\ \midrule
\multirow{6}{*}{Ivanova}   & Origin    & 25.39 $\pm$ 6.20  & 1.46 $\pm$ 0.28 & -                  & -       & -                 & -                     & 0.448 $\pm$ 0.056          & -                \\
                           & Pitch     & 24.83 $\pm$ 6.03  & 1.26 $\pm$ 0.06 & 0.834 $\pm$ 0.143  & 0.265   & 0.964 $\pm$ 0.042 & 0.367 (0.260 - 0.534) & \textbf{0.396 $\pm$ 0.063} & \textbf{0.015}   \\
                           & Formant$_\text{LPC}$   & 22.94 $\pm$ 5.71  & 1.25 $\pm$ 0.05 & 0.625 $\pm$ 0.352  & 0.336   & 0.910 $\pm$ 0.058 & 0.400 (0.267 - 0.532) & 0.503 $\pm$ 0.041          & 0.651            \\
                           & Formant$_\text{spec}$   & 23.71 $\pm$ 5.90  & 1.26 $\pm$ 0.07 & 0.822 $\pm$ 0.241  & 0.196   & 0.968 $\pm$ 0.041 & 0.417 (0.267 - 0.536) & 0.421 $\pm$ 0.023          & 0.163            \\
                           & Neural VC    & 40.57 $\pm$ 2.89  & 1.72 $\pm$ 0.19 & 0.148 $\pm$ 0.221  & 0.327   & 0.915 $\pm$ 0.066 & 0.433 (0.318 - 0.584) & 0.399 $\pm$ 0.042          & 0.136            \\
                           & Resyn   & 38.37 $\pm$ 3.48  & 3.22 $\pm$ 0.24 & 0.036 $\pm$ 0.143  & 0.099   & 0.911 $\pm$ 0.052 & 0.450 (0.297 - 0.553) & \textbf{0.396 $\pm$ 0.045} & $\mathbf{1.57\times 10^{-3}}$ \\ \midrule
\multirow{6}{*}{NeuroVoz}  & Origin    & 18.44 $\pm$ 11.70 & 2.85 $\pm$ 0.67 & -                  & -       & -                 & -                     & 0.862 $\pm$ 0.077             & -                \\
                           & Pitch     & 18.30 $\pm$ 11.48 & 1.38 $\pm$ 0.14 & 0.925 $\pm$ 0.219  & 0.218   & 0.950 $\pm$ 0.059 & 0.435 (0.420 - 0.451) & 0.834 $\pm$ 0.053             & 0.227            \\
                           & Formant$_\text{LPC}$   & 15.53 $\pm$ 9.35  & 1.67 $\pm$ 0.34 & 0.832 $\pm$ 0.366  & 0.184   & 0.932 $\pm$ 0.067 & 0.271 (0.255 - 0.284) & 0.850 $\pm$ 0.075             & 0.500            \\
                           & Formant$_\text{spec}$   & 17.22 $\pm$ 10.76 & 1.46 $\pm$ 0.25 & 0.919 $\pm$ 0.210  & 0.128   & 0.969 $\pm$ 0.050 & 0.226 (0.214 - 0.242) & 0.841 $\pm$ 0.048             & 0.377            \\
                           & Neural VC    & 30.49 $\pm$ 13.78 & 2.20 $\pm$ 0.62 & 0.315 $\pm$ 0.509  & 0.230   & 0.913 $\pm$ 0.070 & 0.481 (0.466 - 0.497) & 0.843 $\pm$ 0.060             & 0.412            \\
                           & Resyn   & 38.22 $\pm$ 13.68 & 3.58 $\pm$ 0.59 & 0.182 $\pm$ 0.398  & 0.173   & 0.908 $\pm$ 0.078 & 0.486 (0.470 - 0.503) & \textbf{0.515 $\pm$ 0.112 }    & $\mathbf{1.18\times 10^{-7}}$ \\ \midrule
\multirow{6}{*}{ENNI}      & Origin    & 28.39 $\pm$ 5.30  & 1.54 $\pm$ 0.29 & -                  & -       & -                 & -                     & 0.849 $\pm$ 0.059           & -                \\
                           & Pitch     & 26.20 $\pm$ 4.96  & 1.25 $\pm$ 0.03 & 0.630 $\pm$ 0.231  & 0.287   & 0.939 $\pm$ 0.069 & 0.467 (0.429 - 0.506) & 0.831 $\pm$ 0.048           & 0.900            \\
                           & Formant$_\text{LPC}$   & 32.25 $\pm$ 5.07  & 1.25 $\pm$ 0.03 & 0.485 $\pm$ 0.300  & 0.172   & 0.943 $\pm$ 0.058 & 0.229 (0.196 - 0.260) & 0.824 $\pm$ 0.090           & 0.879            \\
                           & Formant$_\text{spec}$   & 27.04 $\pm$ 5.00  & 1.25 $\pm$ 0.06 & 0.826 $\pm$ 0.170  & 0.152   & 0.957 $\pm$ 0.052 & 0.256 (0.222 - 0.289) & 0.822 $\pm$ 0.056           & 0.791            \\
                           & Neural VC    & 39.71 $\pm$ 1.51  & 1.79 $\pm$ 0.23 & 0.134 $\pm$ 0.223  & 0.239   & 0.942 $\pm$ 0.046 & 0.489 (0.449 - 0.523) & 0.810 $\pm$ 0.053           & 0.114            \\
                           & Resyn   & 38.00 $\pm$ 5.49  & 4.28 $\pm$ 0.21 & -0.012 $\pm$ 0.128 & 0.113   & 0.927 $\pm$ 0.052 & 0.500 (0.303 - 0.833) & 0.754 $\pm$ 0.055           & 0.184            \\ \bottomrule
                        
\end{tabular}
}
\end{sidewaystableplain}

\begin{sidewaystableplain}[h]
\centering
\caption{\textbf{$|$ Results on voice disorder datasets.}}
\label{tab: result3}
\resizebox{1\linewidth}{!}{
\begin{tabular}{@{}cccccccccc@{}}
\toprule
Dataset                                    & Anonymise & SNR              & MOS             & PCC$_\text{F0}$   & CER/WER & Emo               & EER                   & BACC                           & p-value           \\ \midrule
\multirow{6}{*}{Saarbrücken—HD}            & Origin    & 25.10 $\pm$ 5.98 & 3.29 $\pm$ 0.45 & -                 & -       & -                 & -                     & 0.743  $\pm$ 0.013          & -                 \\
                                           & Pitch     & 24.42 $\pm$ 6.40 & 1.55 $\pm$ 0.23 & 0.966 $\pm$ 0.103 & 0.005   & 0.949 $\pm$ 0.055 & 0.422 (0.387 - 0.461) & 0.717  $\pm$ 0.017 & 0.136    \\
                                           & Formant$_\text{LPC}$   & 22.47 $\pm$ 4.94 & 2.01 $\pm$ 0.41 & 0.780 $\pm$ 0.425 & 0.029   & 0.908 $\pm$ 0.064 & 0.258 (0.231 - 0.292) & 0.764  $\pm$ 0.011 & 0.840    \\
                                           & Formant$_\text{spec}$   & 22.98 $\pm$ 5.73 & 1.92 $\pm$ 0.47 & 0.960 $\pm$ 0.134 & 0.010   & 0.973 $\pm$ 0.039 & 0.242 (0.213 - 0.272) & 0.736  $\pm$ 0.009          & 0.417             \\
                                           & Neural VC    & 30.97 $\pm$ 8.25 & 2.83 $\pm$ 0.35 & 0.443 $\pm$ 0.440 & 0.022   & 0.922 $\pm$ 0.072 & 0.510 (0.478 - 0.546) & \textbf{0.684  $\pm$ 0.021 } & \textbf{0.011}    \\
                                           & Resyn   & 48.41 $\pm$ 7.91 & 3.20 $\pm$ 0.24 & 0.068 $\pm$ 0.457 & 0.015   & 0.913 $\pm$ 0.070 & 0.517 (0.478 - 0.550) & \textbf{0.486  $\pm$ 0.013 } & $\mathbf{3.73\times 10^{-14}}$ \\ \midrule
\multirow{6}{*}{Saarbrücken—Laryngitis}    & Origin    & 24.93 $\pm$ 6.15 & 3.19 $\pm$ 0.50 & -                 & -       & -                 & -                     & 0.807  $\pm$ 0.031          & -                 \\
                                           & Pitch     & 24.25 $\pm$ 6.50 & 1.53 $\pm$ 0.21 & 0.966 $\pm$ 0.108 & 0.027   & 0.943 $\pm$ 0.057 & 0.438 (0.392 - 0.478) & 0.814  $\pm$ 0.042          & 0.686             \\
                                           & Formant$_\text{LPC}$   & 22.31 $\pm$ 5.04 & 1.94 $\pm$ 0.41 & 0.796 $\pm$ 0.451 & 0.029   & 0.908 $\pm$ 0.064 & 0.224 (0.186 - 0.259) & 0.771  $\pm$ 0.057          & 0.092             \\
                                           & Formant$_\text{spec}$   & 22.85 $\pm$ 5.89 & 1.94 $\pm$ 0.50 & 0.960 $\pm$ 0.131 & 0.037   & 0.969 $\pm$ 0.040 & 0.221 (0.183 - 0.250) & 0.789  $\pm$ 0.026          & 0.271             \\
                                           & Neural VC    & 30.52 $\pm$ 7.65 & 2.79 $\pm$ 0.37 & 0.463 $\pm$ 0.426 & 0.043   & 0.907 $\pm$ 0.069 & 0.460 (0.422 - 0.511) & \textbf{0.739  $\pm$ 0.068 } & \textbf{0.011}    \\
                                           & Resyn   & 48.17 $\pm$ 7.22 & 3.24 $\pm$ 0.26 & 0.057 $\pm$ 0.469 & 0.016   & 0.901 $\pm$ 0.068 & 0.526 (0.480 - 0.566) & \textbf{0.489  $\pm$ 0.069 } & $\mathbf{3.08\times 10^{-14}}$ \\ \midrule
\multirow{6}{*}{Saarbrücken—Rekurrensparese} & Origin    & 24.74 $\pm$ 5.70 & 3.04 $\pm$ 0.66 & -                 & -       & -                 & -                     & 0.901  $\pm$ 0.012          & -                 \\
                                           & Pitch     & 24.18 $\pm$ 6.14 & 1.52 $\pm$ 0.21 & 0.952 $\pm$ 0.157 & 0.025   & 0.951 $\pm$ 0.054 & 0.397 (0.358 - 0.427) & \textbf{0.870  $\pm$ 0.020 } & \textbf{0.020}    \\
                                           & Formant$_\text{LPC}$   & 22.29 $\pm$ 4.83 & 1.88 $\pm$ 0.41 & 0.757 $\pm$ 0.440 & 0.037   & 0.911 $\pm$ 0.065 & 0.234 (0.207 - 0.263) & 0.894  $\pm$ 0.028          & 0.380             \\
                                           & Formant$_\text{spec}$   & 22.84 $\pm$ 5.51 & 1.84 $\pm$ 0.48 & 0.955 $\pm$ 0.163 & 0.018   & 0.975 $\pm$ 0.038 & 0.200 (0.173 - 0.229) & \textbf{0.877  $\pm$ 0.021 } & \textbf{0.049}    \\
                                           & Neural VC    & 30.94 $\pm$ 7.77 & 2.74 $\pm$ 0.36 & 0.409 $\pm$ 0.422 & 0.025   & 0.916 $\pm$ 0.073 & 0.500 (0.465 - 0.535) & \textbf{0.830  $\pm$ 0.023 } & $\mathbf{8.81\times 10^{-5}}$ \\
                                           & Resyn   & 48.19 $\pm$ 7.09 & 3.22 $\pm$ 0.26 & 0.017 $\pm$ 0.471 & 0.005   & 0.909 $\pm$ 0.069 & 0.493 (0.456 - 0.526) & \textbf{0.473  $\pm$ 0.025 } & $\mathbf{7.10\times 10^{-14}}$ \\ \midrule
\multirow{6}{*}{PUMCH}                     & Origin    & 23.17 $\pm$ 7.81 & 1.32 $\pm$ 0.07 & -                 & -       & -                 & -                     & 0.699 $\pm$ 0.090          & -                 \\
                                           & Pitch     & 22.62 $\pm$ 7.27 & 1.37 $\pm$ 0.12 & 0.866 $\pm$ 0.158 & 0.192   & 0.951 $\pm$ 0.063 & 0.349 (0.324 - 0.379) & 0.644 $\pm$ 0.033          & 0.058             \\
                                           & Formant$_\text{LPC}$   & 24.55 $\pm$ 7.23 & 1.25 $\pm$ 0.02 & 0.693 $\pm$ 0.327 & 0.154   & 0.954 $\pm$ 0.056 & 0.214 (0.191 - 0.237) & 0.691 $\pm$ 0.032          & 0.445             \\
                                           & Formant$_\text{spec}$   & 22.22 $\pm$ 7.36 & 1.24 $\pm$ 0.02 & 0.934 $\pm$ 0.099 & 0.115   & 0.967 $\pm$ 0.048 & 0.119 (0.099 - 0.140) & 0.648 $\pm$ 0.066          & 0.055             \\
                                           & Neural VC    & 41.36 $\pm$ 1.53 & 1.87 $\pm$ 0.18 & 0.214 $\pm$ 0.258 & 0.241   & 0.933 $\pm$ 0.073 & 0.488 (0.465 - 0.521) & \textbf{0.641 $\pm$ 0.053} & \textbf{0.045}    \\
                                           & Resyn   & 42.12 $\pm$ 13.00 & 2.49 $\pm$ 0.53 & 0.004 $\pm$ 0.199 & 0.045   & 0.935 $\pm$ 0.068 & 0.440 (0.415 - 0.474) & \textbf{0.526 $\pm$ 0.070} & $\mathbf{1.35\times 10^{-5}}$  \\ \bottomrule
                            
\end{tabular}
}
\end{sidewaystableplain}

\begin{figure}[h]
    \centering
    \includegraphics[width=0.9\linewidth]{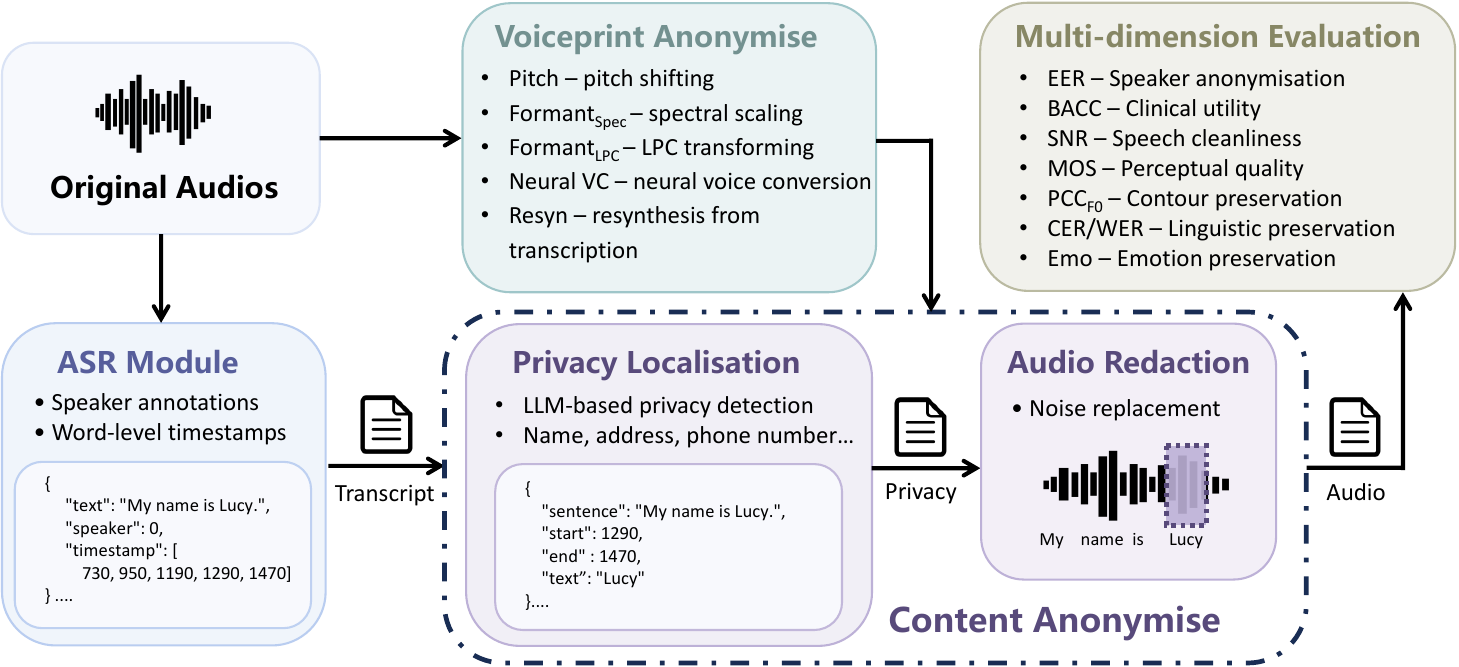}
    \caption{\textbf{$|$ Detailed workflow of CASA anonymisation.} Original recordings are transcribed using an ASR module to obtain transcripts with speaker annotations and word-level timestamps. Available voiceprint anonymisation methods include pitch shifting (\texttt{Pitch}), spectral formant scaling (\texttt{Formant$_\text{spec}$}), LPC-based formant transformation (\texttt{Formant$_\text{LPC}$}), neural voice conversion (\texttt{Neural VC}) and transcription–resynthesis (\texttt{Resyn}). For content anonymisation, a locally deployed LLM identifies sensitive spans in the transcripts and produces annotations to guide redaction. For \texttt{Resyn}, which doesn't preserve temporal alignment, sensitive content is masked in the transcript before synthesis; while for other methods, the corresponding speech segments are replaced with noise. The resulting recordings are evaluated for speaker privacy (EER), downstream clinical utility (BACC), signal cleanliness (SNR), perceptual quality (MOS), pitch-contour preservation (PCC$_\text{F0}$), linguistic preservation (CER/WER) and emotion preservation (Emo).}
    \label{fig:workflow}
\end{figure}

\begin{figure}[h]
    \centering
    \includegraphics[width=0.6\linewidth]{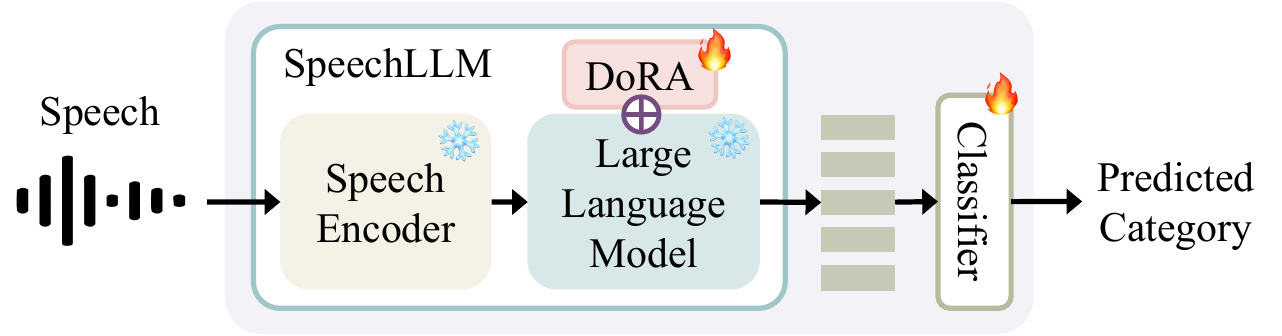}
    \caption{\textbf{$|$ Downstream model structure.}}
    \label{fig:downstream_model}
\end{figure}

\begin{figure}[h]
    \centering
    \includegraphics[width=1.0\linewidth]{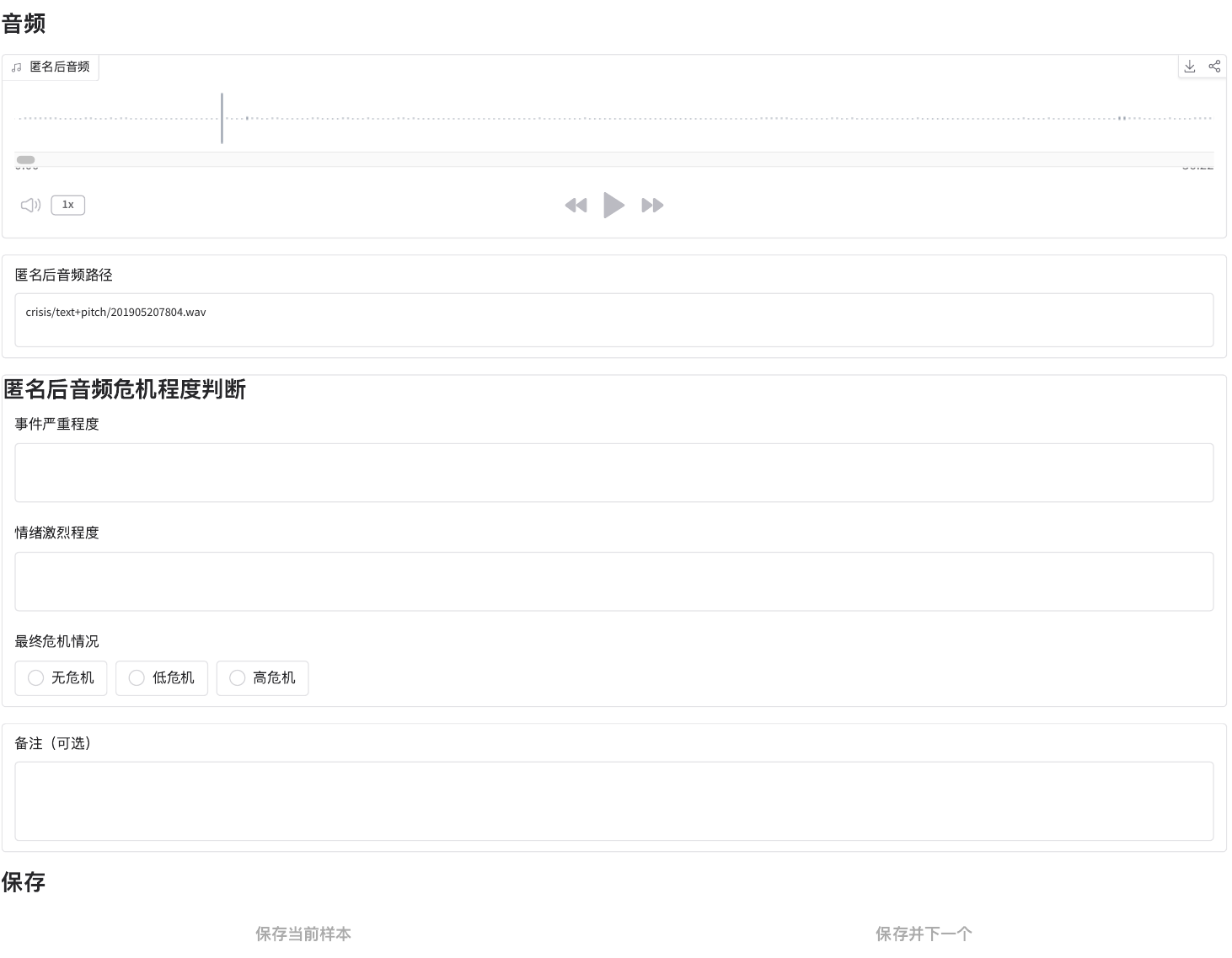}
    \caption{\text{$|$ Web interface for crisis relabelling.}}
    \label{fig:web_crisis_util}
\end{figure}

\begin{figure}
    \centering
    \includegraphics[width=1.0\linewidth]{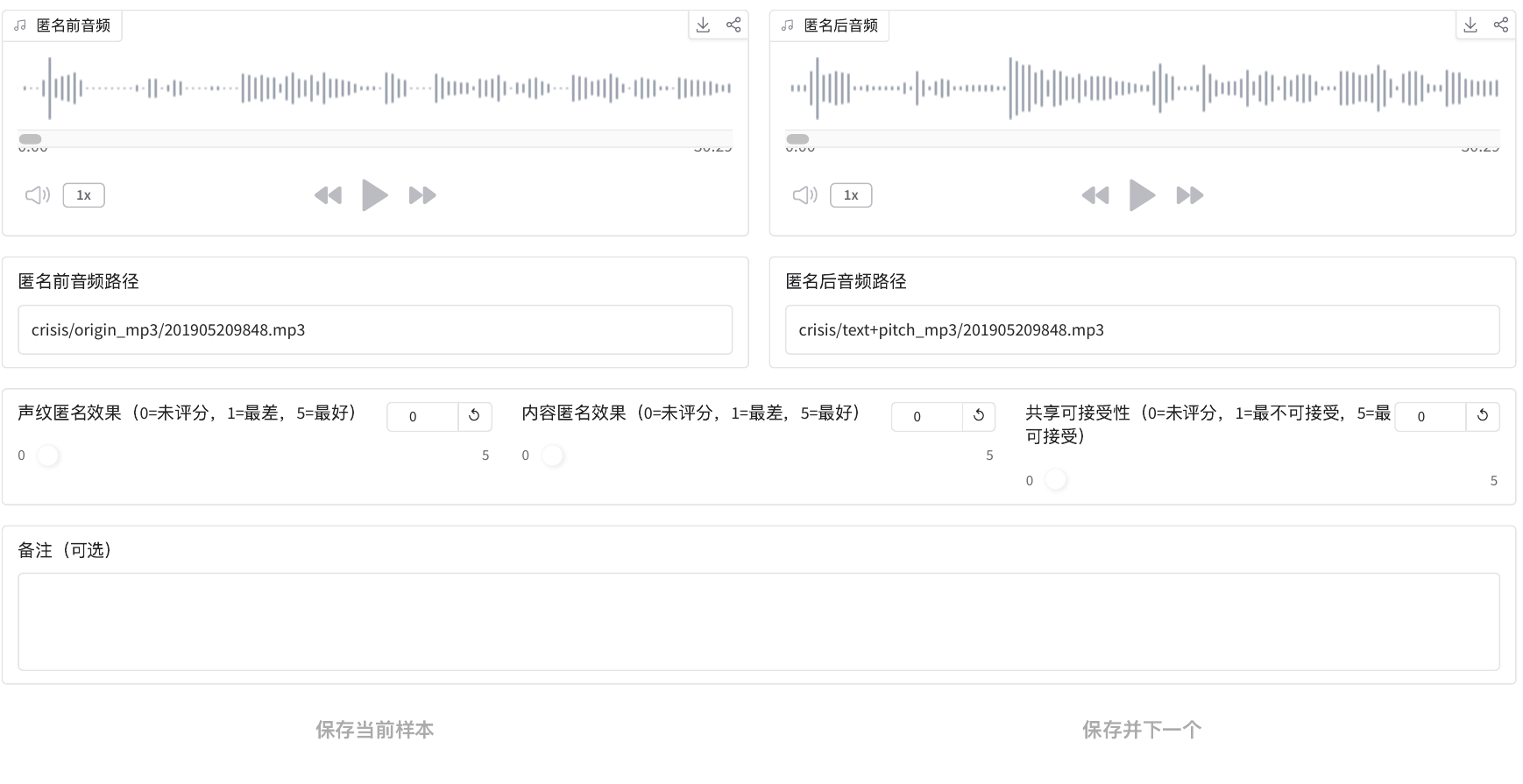}
    \caption{\textbf{$|$ Web interface for crisis anonymisation rating.}}
    \label{fig:web_crisis_anony}
\end{figure}

\begin{figure}
    \centering
    \includegraphics[width=1.0\linewidth]{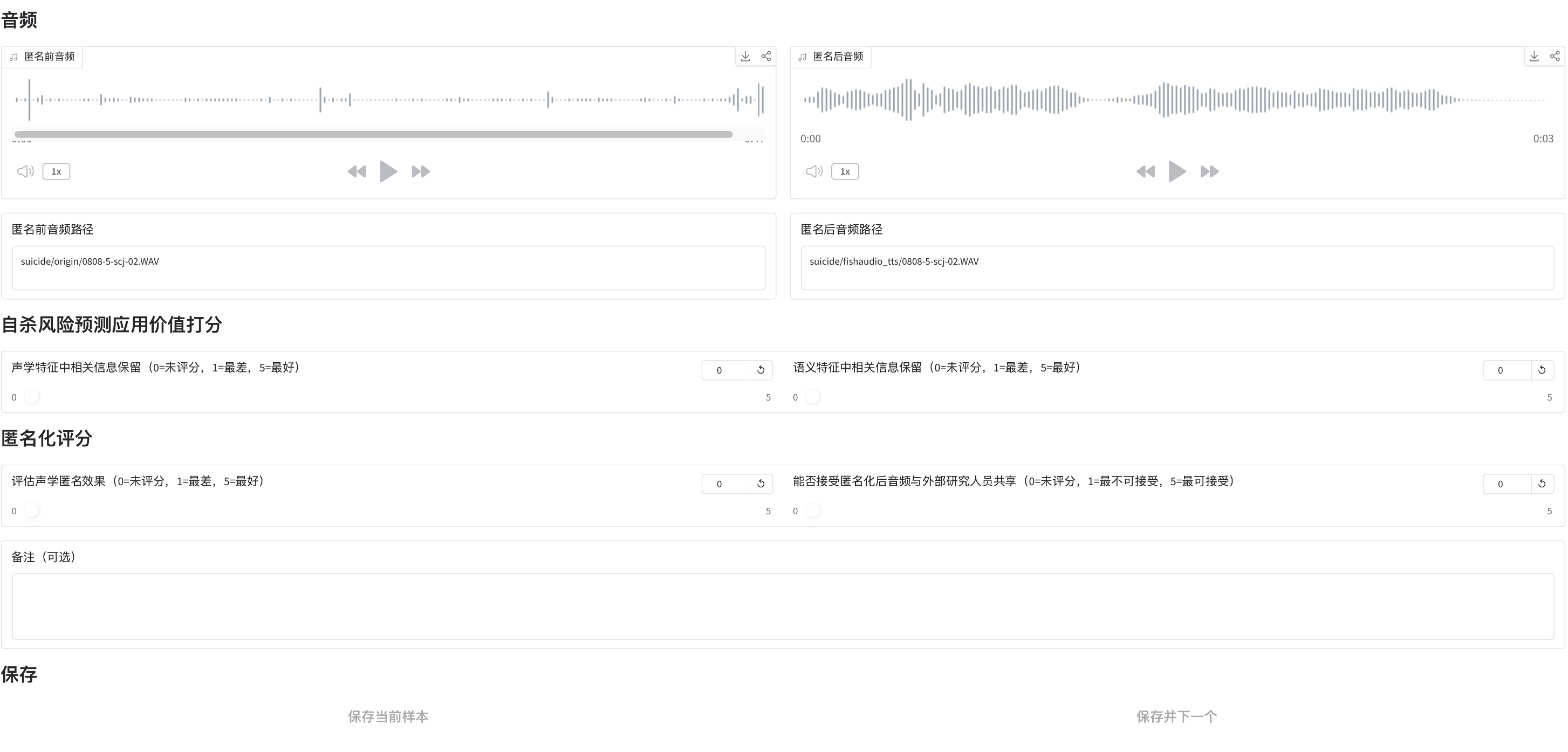}
    \caption{\textbf{$|$ Web interface for Suicide-YF scoring.}}
    \label{fig:web_suicide}
\end{figure}

\begin{figure}
    \centering
    \includegraphics[width=1.0\linewidth]{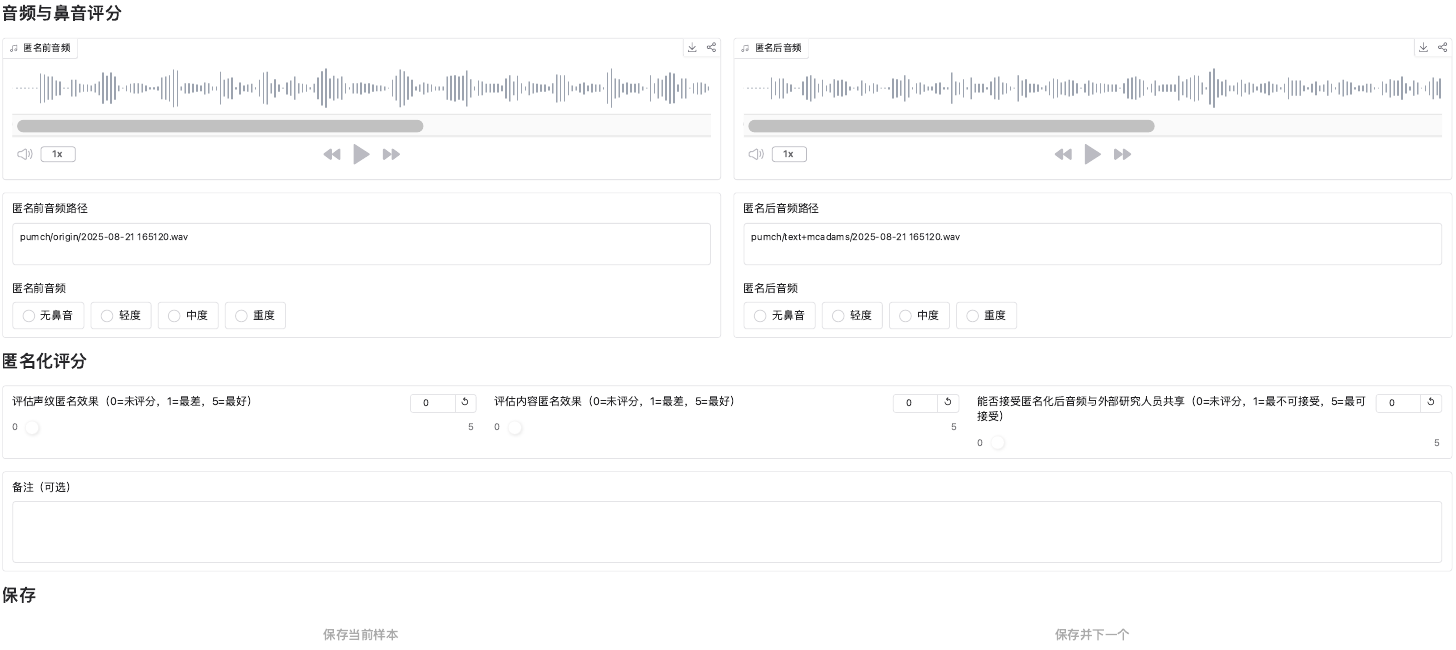}
    \caption{\textbf{$|$ Web interface for PUMCH evaluation.}}
    \label{fig:web_pumch}
\end{figure}

\end{appendices}



\end{document}